%% file: main.tex
\documentclass[conference]{IEEEtran}
\IEEEoverridecommandlockouts
\usepackage{multirow}
\usepackage{multicol}
\usepackage{adjustbox}
\usepackage{amsmath}
\usepackage{mathtools}
\usepackage{subcaption}
\usepackage{paralist}
\usepackage{todonotes}
\usepackage{booktabs}

\def\BibTeX{{\rm B\kern-.05em{\sc i\kern-.025em b}\kern-.08em
    T\kern-.1667em\lower.7ex\hbox{E}\kern-.125emX}}
    
\begin{document}

\title{Membership Privacy Evaluation\\ in Deep Spiking Neural Networks} %*\\
% {\footnotesize \textsuperscript{*}Note: Sub-titles are not captured for https://ieeexplore.ieee.org  and

% \author{\IEEEauthorblockN{Anonymous Authors}}

\author{\IEEEauthorblockN{Jiaxin Li\textsuperscript{*}}
\IEEEauthorblockA{%\textit{dept. name of organization (of Aff.)} \\
\textit{University of Padua}\\
Padova, Italy \\
jiaxin.li@studenti.unipd.it}
\and
\IEEEauthorblockN{Gorka Abad}
\IEEEauthorblockA{%\textit{dept. name of organization (of Aff.)} \\
\textit{Radboud University}\\
Nijmegen, Netherlands\\
abad.gorka@ru.nl}
\and
\IEEEauthorblockN{Stjepan Picek}
\IEEEauthorblockA{%\textit{dept. name of organization (of Aff.)} \\
\textit{Radboud University}\\
Nijmegen, Netherlands\\
stjepan.picek@ru.nl}
\and
\IEEEauthorblockN{Mauro Conti}
\IEEEauthorblockA{%\textit{dept. name of organization (of Aff.)} \\
\textit{University of Padua}\\
Padova, Italy \\
conti@math.unipd.it}
\thanks{\textsuperscript{*}Corresponding author.}
}
% \and
% \and

\maketitle

% \begin{abstract}
% This document is a model and instructions for \LaTeX.
% This and the IEEEtran.cls file define the components of your paper [title, text, heads, etc.]. *CRITICAL: Do Not Use Symbols, Special Characters, Footnotes, 
% or Math in Paper Title or Abstract.
% \end{abstract}

\input{sections/abstract}


\input{sections/introduction}

\input{sections/background}

\input{sections/our_attack}

\input{sections/experiments}

\input{sections/results_and_discussion}

\input{sections/related_works}

\input{sections/conclusion}

% \section*{Acknowledgment}
% This research was supported by the Chinese Scholarship Council (CSC).


% \section*{References}

\bibliographystyle{IEEEtran}
\bibliography{paper}

\appendix

\input{sections/appendix}

\end{document}

%% file: sections/abstract.tex
\begin{abstract}
Artificial Neural Networks (ANNs), commonly mimicking neurons with non-linear functions to output floating-point numbers, consistently receive the same signals of a data point during its forward time. Unlike ANNs, Spiking Neural Networks (SNNs) get various input signals in the forward time of a data point and simulate neurons in a biologically plausible way, i.e., producing a spike (a binary value) if the accumulated membrane potential of a neuron is larger than a threshold. Even though ANNs have achieved remarkable success in multiple tasks, e.g., face recognition and object detection, SNNs have recently obtained attention due to their low power consumption, fast inference, and event-driven properties. 
While privacy threats against ANNs are widely explored, much less work has been done on SNNs. For instance, it is well-known that ANNs are vulnerable to the Membership Inference Attack (MIA), but whether the same applies to SNNs is not explored.

In this paper, we evaluate the membership privacy of SNNs by considering eight MIAs, seven of which are inspired by MIAs against ANNs. Our evaluation results show that SNNs are more vulnerable (maximum 10\% higher in terms of balanced attack accuracy) than ANNs when both are trained with neuromorphic datasets (with time dimension). On the other hand, when training ANNs or SNNs with static datasets (without time dimension), the vulnerability depends on the dataset used. 
If we convert ANNs trained with static datasets to SNNs, the accuracy of MIAs drops (maximum 11.5\% with a reduction of 7.6\% on the test accuracy of the target model). 
Next, we explore the impact factors of MIAs on SNNs by conducting a hyperparameter study. Finally, we show that the basic data augmentation method for static data and two recent data augmentation methods for neuromorphic data can considerably (maximum reduction of 25.7\%) decrease MIAs' performance on SNNs. Regardless, the accuracy of MIAs could still be between 51.7\% and 66.4\% with data augmentation, indicating data augmentation cannot fully prevent MIAs on SNNs.

\end{abstract}

%% file: sections/introduction.tex
\section{Introduction}
\label{sec:introduction}

% \todo{g: im missing an initial paragraph with a broader intro to SNNs or DL in general. It seem like you are directly diving into SNNs with some complex terms in the paragraph straightaway--response: Since we have much content to introduce the ANNs, SNNs, and their application in the first two paragraphs, adding a broader intro might make the intro too long. Hence, I prefer to keep it.}

\begin{figure}
    % \Description[neuron in ANN and SNN]
    \centering
    \begin{subfigure}[b]{0.48\textwidth}
        \centering
        \includegraphics[scale=0.5]{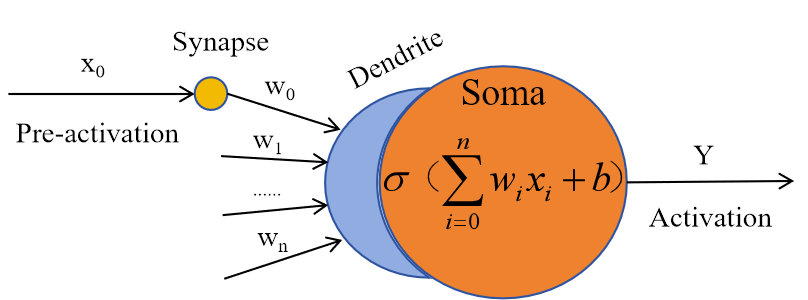}
        \caption{A neuron in ANN.}
        \label{neuron_ANN}
    \end{subfigure}
    \begin{subfigure}[b]{0.48\textwidth}
        \centering
        \includegraphics[scale=0.5]{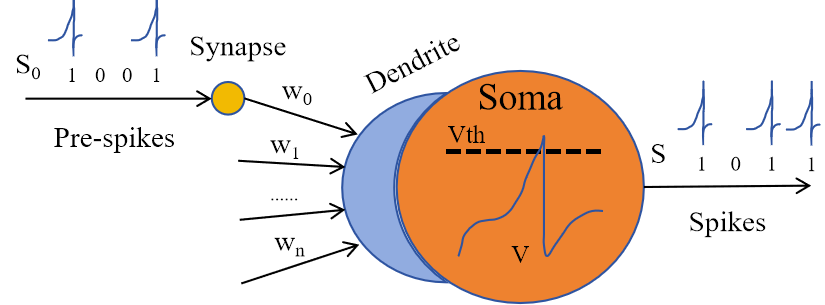}
        \caption{A neuron in SNN.}
        \label{neuron_SNN}
    \end{subfigure}
    \caption{Basic neuron model in ANNs and SNNs~\cite{Deng_2020_Rethink}.}
    \label{neuon_model_of_ANN_and_SNN}
\end{figure}

Artificial Neural Networks model the behavior of a neuron with non-linear functions. As large amounts of data are collected and computing capabilities are enhanced, ANNs, especially deep neural networks~\cite{Goodfellow_2016_deep_learning}, demonstrate an amazing ability to solve real-world tasks like face recognition~\cite{deepface_yaniv} and object detection~\cite{zou2023object}. The third generation of neural network models, Spiking Neural Networks~\cite{pfeiffer2018deep}, mimic the dynamics of a neuron in a way closer to the actual neurons in the brain. Figure~\ref{neuon_model_of_ANN_and_SNN} shows the basic neuron models for ANNs and SNNs. For ANNs, the neuron collects weighted inputs from previous neurons, applies a non-linear function $\sigma$ to the summed input, and continuously outputs an identical activation value if the input is the same over time. For SNNs, the neuron modifies its membrane potential $V$ (and the rate of change) according to binary spikes from previous neurons in the current time step. It outputs a spike when the membrane potential exceeds a threshold $V_{th}$. If there are no spikes from previous neurons, the membrane potential will gradually reset to a low value to save energy. The neuron remains inactive until a new spike is received, which improves power consumption. Due to those properties and the development of neuromorphic devices (e.g., TrueNorth from IBM~\cite{TrueNorth_IBM} and Loihi from Intel~\cite{Loihi_Inter}), SNNs gained much attention in scenarios like the autonomous operation of the vehicle~\cite{Cheng_Autonomous_Driving_2020}, industrial fault diagnosis~\cite{wang2023brain}, and healthcare diagnosis with biomedical signals~\cite{choi2024spiking,xiaoxue2023review}.

Since the initially proposed SNNs could be hard to train and usually perform worse than ANNs~\cite{pfeiffer2018deep, tavanaei2019deep, Deng_2020_Rethink}, many works follow lessons from ANNs to improve the performance of SNNs. Specifically, directly converting trained ANNs to SNNs~\cite{cao2015spiking,diehl2015fast,rueckauer2017conversion}, increasing the depth of the model~\cite{Sengupta_2019_Deeper_SNN,fang2021deep}, replacing the non-differential threshold function with the differential surrogate function~\cite{Hunsberger_2015_SNN_with_LIF}, and training SNNs with backpropagation~\cite{Lee_2016_SNN_Backpropagation}. Moreover, we can distinguish shallow and deep SNNs according to the number of layers within the structure of SNNs following the previous work~\cite{Sengupta_2019_Deeper_SNN,fang2021deep}. 
As deep SNNs could perform better than shallow SNNs~\cite{Sengupta_2019_Deeper_SNN,fang2021deep} and since they are used in real-world tasks~\cite{Cheng_Autonomous_Driving_2020,lei2024dt}, our work focuses on deep SNNs trained with backpropagation or converted from ANNs.

%care and respect customers' privacy

Membership Inference Attacks, inferring whether a data point is in the training data of the target model, attract much attention in ANN research and practice, as indicated by recent laws like GDPR~\cite{GDPR_citation} and CCPA~\cite{CCPA_citation}. 
Applying SNNs in real-world tasks raises similar membership privacy concerns as in the case of ANNs. 
More precisely, being in the training data of SNNs could disclose the private information of users. For example, if the user's data is in the training data of the deep SNN trained for classification of Alzheimer's disease~\cite{turkson2021classification}, we could infer that this user suffered from this disease if we know the data of this user is in the training data via MIA. 
As far as we know, no studies have evaluated the membership privacy of deep SNNs. As such, we address this gap in our work.

Intuitively, MIAs on SNNs could be more challenging than MIAs on ANNs, considering a data point of the neuromorphic dataset consists of multiple frames, each of which is the accumulation of events that happen over a period. Let us consider the image classification task.
A data point in the static dataset only contains one static image. A data point with multiple frames has more neighboring data points than the one with one static image. More neighboring data points make it harder to predict the existence of a specific data point, as the inclusion of neighboring data points could disturb the prediction of the MIA in this particular data point~\cite{Long_2020_Pragmatic}. 
To solve this problem, we utilize spiking neurons' fire rate and membrane potential as signals of MIAs since they are cumulative results of multiple frames and could represent the prediction situation of multiple frames. 
In commonly used rate coding SNNs~\cite{auge2021survey}, the fire rate is a similar indicator to the confidence score to show the confidence of the prediction due to the alignment of Mean Squared Error (MSE) during training. Hence, we utilize previous strategies~\cite{shokri_membership_2017,salem_ml-leaks_2018,carlini_membership_2021} in MIAs with confidence scores to MIAs with fire rates.

The membership evaluation results show that SNNs are more vulnerable than ANNs when training ANNs and SNNs with neuromorphic datasets. On the other hand, when working with static datasets, the vulnerability comparison between ANNs and SNNs depends on specific datasets. 
The conversion from ANNs to SNNs maximally reduces 11.5\% on the performance of MIAs with a drop of 7.6\% on the original classification task.
In the hyperparameter study, we show that the ATan function and the Leaky Integrate and Fire (LIF) neuron bring a high classification accuracy with a large generalization gap, leading to MIAs' high attack accuracy. Next, the choices of Adam (with a learning rate of 0.001) and SGD (with a learning rate of 0.1) are suitable for training SNNs. Moreover, increasing the number of time steps will slightly increase the vulnerability under MIAs.
Finally, the basic data augmentation method for static data and two recent data augmentation methods for neuromorphic data could reduce MIAs' performance on SNNs by up to 25.7\%. Unfortunately, the accuracy of MIAs could still be from 51.7\% to 66.4\% even when applied data augmentation.
% In the ablation study, attempts that enlarge the generalization gap of the target model (training accuracy minus test accuracy) will usually increase the performance of MIAs. After applying two data augmentation mechanisms to neuromorphic datasets, we observe they can reduce the generalization gap (even keep the test accuracy of the target model) to decrease the performance of MIAs.\todo{g: this is expected, right?}

Our main contributions are:
\begin{compactenum}
    \item To the best of our knowledge, we are the first to evaluate the membership privacy of SNNs and compare their vulnerability to ANNs. We experiment with eight MIAs, six datasets, and three model structures, which lay the foundation for future work on the privacy of SNNs.
    \item To understand MIAs on SNNs better, we investigate influential factors of MIAs during the training of SNNs in the hyperparameter study.
    % \item We utilize two data augmentation mechanisms to improve the generalization of target models and reduce the threat of MIAs.
    \item We apply data augmentation to defend against MIAs on SNNs. There, we show that the basic augmentation method for static datasets and two recent augmentation mechanisms for neuromorphic datasets can reduce but cannot completely prevent MIAs on SNNs.
\end{compactenum}
Following the paper's acceptance, we will make our source code available to the community.

%% file: sections/background.tex
\section{Background}
In Section~\ref{background_SNN}, we introduce SNNs and three main methods to build SNNs. Next, we give a simple explanation of the pipeline of MIAs in Section~\ref{background_MIA}.

\subsection{Spiking Neural Network}
\label{background_SNN}

% \todo{g: overall i see this entire section too complex. I would rather explain less things that are crucial to understand the attack. The rest either you mention it or skip it. Because its too complex for the reader to keep track of everything and distinguish what is important.}

% The transmitted information in the human brain is a sequence of pulse data with only two states: On and Off. The animals use the temporal and quantitative differences of signals received by their two ears to recognize sound direction, which means they use both quantity and time signals. Kuroyanagi et al.~\cite{Kuroyanagi_1993_AuditoryNN} proposed a pulse neuron model to transmit pulse trains\todo{?} from multiple synapses to the internal potential of the neuron via weighting and decaying pulse trains over time, which is the initial prototype of the spiking neural network (SNN) \todo{g: already introduced}. It corresponds to the physiological factors that the brain communicates with pulse trains\todo{what trains? trains are means of transportation going on tracks}, and the neurons are connected with hundreds of synapses. 

%the mechanism of %In the brain, the communication between neurons is done by broadcasting a sequence of action potentials, also known as spike trains, to downstream neurons. The SNN is composed of neurons connected via synapses. The synaptic weights reflect the strength of the neurons' connection, which sculpts the functions of SNNs. 

%Neuroscience and Artificial Intelligence (AI) are two interconnected fields~\cite{hassabis2017neuroscience}. 
An SNN is an application of the biological neuron into AI for efficient and low-cost computation. There are three main methods to train SNNs.

\textbf{(1) Spike-Timing Dependent Plasticity (STDP).} For a pair of presynaptic and postsynaptic neurons, the synaptic weight of the synapse between those two neurons alters according to the arrival time of spikes from the presynaptic neuron and the firing time of the postsynaptic neuron. As an example, we consider a neuron ``A'' and a neuron ``B'' to be connected. The faster ``A'' fires, the stronger the connection with ``B'', and vice versa; when ``A'' does not fire as much, the connection is weakened. The weight will increase if the presynaptic neuron fires early and the postsynaptic neuron fires later. The weight will decrease if the presynaptic neuron fires later while the postsynaptic neuron fires early.

However, we do not consider this strategy because STDP is suitable for shallow SNNs. For deep SNNs, finding suitable hyperparameters and training with STDP is difficult, as mentioned in the previous work~\cite{polap2022heuristic}. Our focus is deep SNNs, and we obtain deep SNNs trained via backpropagation or converted from ANNs.

% Following previous works~\cite{morrison2008phenomenological,sjostrom2010spike}, we formulate the total change ($\Delta w_{ij}$) of the synaptic weight ($w_{ij}$) of the synapse between the presynaptic neuron $j$ and postsynaptic neuron $i$ as Eq.~\eqref{total_change_of_weight}. $t^{pre}$ (e.g. 0, 3, 4, $\dots$, ms) and $t^{post}$ are fire times of neurons $j$ and $i$ till current time $t$. $W(x)$ denotes one of the STDP functions, and a popular choice is represented in Eq.~\eqref{STDP_function}. $F_{+}$ and $F_{-}$ are related to current weight $w_{ij}$. $\tau_{+}$ and $\tau_{-}$ are time constants.
% % $t_m^{pre}$ and $t_n^{post}$ are $m$-th and $n$-th fire times of $t^{pre}$ and $t^{post}$ separately.

% \begin{equation}
%     \label{total_change_of_weight}
%     % \Delta w_{ij}=\sum_{m=1}^{|t^{pre}|}\sum_{n=1}^{|t^{post}|}W(t_n^{post}-t_m^{pre}).
%     \Delta w_{ij}=\sum_{t_s^{pre} \in t^{pre}}\sum_{t_{s}^{post} \in t^{post}}W(t_s^{post}-t_s^{pre}).
% \end{equation}

% \begin{equation}
%     \label{STDP_function}
%     W(x)= \left\{
%     \begin{array}{rcl}
%          F_{+}(w_{ij})\exp{(-x/\tau_{+})} & \mbox{for} & x>0\\
%          -F_{-}(w_{ij})\exp{(x/\tau_{-})} & \mbox{for} & x<0
%     \end{array}\right.
% \end{equation}
% \todo{g: whats exp?}

%pointed out conversion challenges: negative output values in ANN layers are difficult to represent in SNNs, biases in convolution layers are problematic, and max-pooling implementation requires two-layer networks (a lateral inhibition followed by pooling over these small image regions) \todo{g: why?}. They during the training of the ANN 

\textbf{(2) Conversion from ANN.} Under this strategy, a pre-trained ANN is converted to an SNN by replacing the ReLU activation layers (that are not used in SNNs) with spiking neurons and adding scaling operations like weight normalization and threshold balancing~\cite{cao2015spiking,diehl2015fast, Hunsberger_2015_SNN_with_LIF,rueckauer2016theory,rueckauer2017conversion}. 
Cao et al.~\cite{cao2015spiking} proposed using the absolute values of negative activations, changing Tanh to ReLU, removing biases, and using spatial linear subsampling instead of max-pooling. Diehl et al.~\cite{diehl2015fast} further analyzed performance loss and suggested weight normalization methods to address the over- and under-activation of spiking neurons.
Instead of applying the ReLU activation function while training an ANN, Hunsberger et al.~\cite{Hunsberger_2015_SNN_with_LIF} used a modified non-linearity LIF neuron and injected noise during training, improving the robustness of the converted SNN's approximation errors. 
Rueckauer et al.~\cite{rueckauer2016theory,rueckauer2017conversion} addressed approximation errors between spiking neuron fire rates and ANN activations, which degrade the accuracy of deep models, by resetting potentials through subtraction and retaining biases as constant input currents scaled by maximum ReLU activation.

% They proposed robust normalization using the $p$-th percentile of the activation distribution to discard outliers, boost firing rates, and adjust weights and biases for batch normalization using batch mean, variance, and learnable parameters. Analog inputs are fed directly to the first hidden layer to reduce undersampling, bypassing Poisson firing rates\todo{what are those?}. For spiking softmax, the softmax function is applied to accumulated membrane potentials, generating spikes with Poisson generators or directly from softmax values. They use gating functions for max-pooling to permit spikes only from the maximally firing neuron.

In our experiments to attack converted SNNs with MIAs, we follow previous works~\cite{rueckauer2016theory,rueckauer2017conversion}:
i) reset the potential by subtraction (the membrane potential threshold $V_{th}=1.0$), ii) re-scale weights and biases with the robust normalization ($p=99.9\%$), iii) achieve batch normalization via scaling weights and biases, and iv) directly feed the input to the converted SNN, and pass the output through a softmax function. We do not consider max-pooling because it requires estimating presynaptic firing rates, which is computationally complex~\cite{rueckauer2016theory} and does not give much benefit. Hence, we follow the strategy of average pooling from the work of Dieh et al.~\cite{diehl2015fast}.

% Lee et al.~\cite{Lee_2016_SNN_Backpropagation} treated the membrane potentials of spiking neurons as differentiable signals and considered discontinuities at spike times as noise, making it feasible to utilize the error backpropagation mechanism for direct training of deep SNNs with spike signals and membrane potentials. The key idea of their approach is to generate a continuous and differentiable signal on which SGD can work, using low-pass filtered spiking signals added onto the membrane potential and treating abrupt changes of the membrane potential as noise during error backpropagation. Wu et al.~\cite{Wu_2018_STBP} proposed a spatiotemporal backpropagation algorithm to combine layer-by-layer spatial domain and temporal domain via an iterative update of membrane potential in the LIF model~\cite{gerstner2014neuronal}. Besides, they approximated the derivative of the threshold-based spike with four differentiable curves. Instead of only learning synaptic-related parameters, Fan et al.~\cite{Fan_2021_Learnable_Membrane_Time_Constant} included the membrane time constant as a learnable parameter rather than manually tuning it, which makes the training of SNNs less sensitive to initial values and speeds up learning.

\textbf{(3) Backpropagation-based supervised learning.} Under this strategy, we train SNNs via backpropagation, the same way as training ANNs. Among various neural models~\cite{Izhikevich_2003_simple_SN,borisyuk1997information}, the Leaky Integrate and Fire (LIF) model can mimic the behavior of the biological neuron with a minimum number of circuit elements~\cite{dutta2017leaky} and bring a lower complexity. As LIF is frequently used in previous works~\cite{gerstner2014neuronal, Lee_2016_SNN_Backpropagation, Fan_2021_Learnable_Membrane_Time_Constant}, we utilize the LIF model as well. 
The membrane potential of a spiking neuron $i$ under the LIF model is formulated as given in Eqs.~\eqref{neuronal_dynamics} and~\eqref{pre_input}. In Eq.~\eqref{neuronal_dynamics}, $V_{rest}$ is the membrane potential of a neuron without any input. We set $V_{rest}$ as 0 following the related work~\cite{Fan_2021_Learnable_Membrane_Time_Constant}. $V(t)$ is the membrane potential of neuron $i$ at time $t$. $\tau$ is the membrane time constant. In Eq.~\eqref{pre_input}, the neuron $i$ has $n_i$ presynaptic neurons. For a presynaptic neuron $j$, it has a list of spiking times $t_{j}^{pre}$, $w_{ij}$ is the synaptic weight between neuron $i$ and a presynaptic neuron $j$, and $\delta(t)$ is the Dirac delta function (if $t\neq0$, $\delta(t)=0$. $\int_{-\infty}^{\infty} \delta(t)dt=1$).

\begin{equation}
    \label{neuronal_dynamics}
    \textit{$\tau\frac{dV(t)}{dt}=-(V(t)-V_{rest})+X(t).$}
\end{equation}

\begin{equation}
    \label{pre_input}
    \textit{$X(t)=\sum_{j=1}^{n_i} \sum_{t_{s}^{pre}\in t_{j}^{pre}}w_{ij}\delta(t-t_s^{pre}).$}
\end{equation}

To consider $V(t)$ in discrete time, we obtain Eq.~\eqref{discrete_time_version} after applying the Euler method to Eq.~\eqref{neuronal_dynamics}.
\begin{equation}
    \label{discrete_time_version}
    \textit{$H(t)=V(t-1)+\frac{1}{\tau}(-(V(t-1)-V_{rest})+X(t)).$}
\end{equation}

Apart from the neuronal dynamics, the neuron will elicit a spike $S(t)$ to the subsequent neurons if the membrane potential is larger than a threshold $V_{th}$, which is formulated in Eq.~\eqref{threshold_eq}. After eliciting a spike, the membrane potential $V(t)$ will be reset as $V_{reset}$ as explained in Eq.~\eqref{reset_potential}.

\begin{equation}
    \label{threshold_eq}
    \textit{$S(t)= \left\{
    \begin{array}{rcl}
         1 & \mbox{if} & H(t)\geq V_{th},\\
         0 & \mbox{others}.
    \end{array}\right.$}
\end{equation}

\begin{equation}
    \label{reset_potential}
    \textit{$V(t)=H(t)(1-S(t))+V_{reset}S(t).$}
\end{equation}

To apply backpropagation to the training of SNN, it is a common practice to keep $S(t)$ for the forward pass and replace $S(t)$ with a differentiable surrogate function to calculate the backward gradient of the backpropagation as $S(t)$ is not differentiable~\cite{Neftci_Surrogate_2019}. Among potential surrogate functions, we set Atan as the default one as the performance of the original task is higher than applying other explored surrogate functions, as shown in the hyperparameter study in Section~\ref{ablation_study}. The Atan function is formulated as Eq.~\eqref{Atan_function}, where Eq.~\eqref{Atan_gradient} is its gradient. $\alpha$ is the pre-defined parameter with a default value of 2.0, following the implementation of SpikingJelly~\cite{fang2023spikingjelly}.

\begin{equation}
    \label{Atan_function}
    \textit{$S(t)= \frac{1}{\pi}\arctan(\frac{\pi}{2} \alpha H(t))+\frac{1}{2}.$}
\end{equation}

\begin{equation}
    \label{Atan_gradient}
    \textit{$\frac{\partial S(t)}{\partial H(t)}=\frac{\alpha}{2(1+({\frac{\pi}{2} \alpha H(t))}^{2})}.$}
\end{equation}

In the forward computation of each frame within the model, SNN aligns the spiking outputs to the one-hot encoding representation of the target category to make the spiking neuron belonging to the target category output a spike while other neurons do not. An MSE loss is calculated according to the matching degree of the spiking output and the one-hot encoding representation. The optimization guided by the MSE loss via Adam~\cite{Adam_Diederik_2015} or SGD~\cite{ruder2016overview}, standard optimization algorithms in ANNs, implements the alignment by updating the weights.

\subsection{Membership Inference Attack}
\label{background_MIA}

MIAs try to infer if a data point has been used during a target model's training by analyzing it or by observing its behavior to arbitrary inputs. Since the fist works of MIA in machine learning~\cite{shokri_membership_2017,salem_ml-leaks_2018}, many works have investigated MIAs with different adversary knowledge~\cite{nasr2019comprehensive,sablayrolles_white-box_2019,choquette-choo_label-only_2021,label_only_MIA_Mauro}, various model types~\cite{Privacy_Risk_He,EncoderMI_liu,hayes_logan_2019}, and distinct attack methods~\cite{yeom_privacy_2018,Long_2020_Pragmatic,carlini_membership_2021,Ye_Enhanced_2022,Hui_21_BlindMIA}. Formally, MIA is formulated as a function $A: x, M, \Omega \rightarrow \{0,1\}$ with a data point $x$, the target model $M$, and external knowledge $\Omega$ of the adversary. The output $1$ means that $x$ is a member of $M$'s training data and $0$ otherwise.

%The data used for training the attack model is from the data of the shadow model, a model controlled by the adversary to mimic the behavior of the target model, as the adversary knows the training and test data of the shadow model 

Currently, there are two main strategies for implementing MIAs: classifier-based and threshold-based methods. In classifier-based methods, the adversary trains the attack model, a classifier, to predict whether a data point is a member based on features (e.g., the confidence scores~\cite{shokri_membership_2017}) extracted from this data point from the target model. The adversary trains a shadow model to mimic the target model's behavior. The adversary does not have access to the original training data but to some data with a similar distribution to the original. Then, the adversary constructs an attack model by using a dataset extracted from the features obtained by querying the shadow model.

For the threshold-based methods, the adversary directly compares a data point's metric (e.g., loss~\cite{yeom_privacy_2018}) with a threshold to predict its membership. Usually, this threshold is determined according to the average value of the metric in the training data or is selected as the one that obtains high performance on the metric values of the shadow model.

%% file: sections/our_attack.tex
\section{Membership Privacy Evaluation}

In this part, we define a data point in the neuromorphic dataset and the input and output of the SNN in Section~\ref{mem_definition}. In Section~\ref{mem_threat_model}, we provide the threat model. Section~\ref{evaluation_methodology} explains the methods with detailed metrics to evaluate the membership privacy of SNNs.

\subsection{Definitions}
\label{mem_definition}

A data point in the training data of SNNs differs from a data point in the training data of ANNs. Let us take the image classification task as an example. A data point used for training ANNs is a three-dimensional (RGB) array, each of which is a pixel value (after normalization) ranging from 0 to 1. In rate coding SNNs, a data point is a list of time-series events measuring the brightness change during the relative movement between the object (or its image) and the Dynamic Vision Sensor (DVS) camera. The DVS camera will generate a positive polarity event with coordinate information if the brightness of a pixel increases over a threshold. If the brightness decreases below a threshold, the DVS camera will generate a negative polarity event with coordinate information. Hence, there are two types of events, categorized in two dimensions (positive and negative), in the list of events that belong to a data point. In neuromorphic datasets from SpikeJelly~\cite{fang2023spikingjelly}, a list of time-series events is accumulated into a fixed number (i.e., time steps) of frames for each data point. This indicates a data point of SNNs is a fixed number of frames. 

Instead of inferring the membership of one static image in ANNs, MIAs on SNNs need to consider the membership of multiple static images (frames). A data point with multiple static images has more neighboring data points than one with one static image. More neighboring data points make it harder to predict the existence of a specific data point, as the inclusion of neighboring data points could disturb the prediction of MIA~\cite{Long_2020_Pragmatic}. To solve this problem, we utilize the fire rate of spiking neurons and the membrane potential as the features. The reason for selecting the fire rate is that the fire rate contains the spiking outputs of multiple frames of a data point rather than the spiking output of one step. Hence, the fire rate is suitable for representing the prediction towards multiple frames. 
As for the membrane potential, the forward of one frame in the SNN will modify the membrane potential of neurons in the SNN, and the accumulation of membrane potential on multiple frames represents the prediction situation of multiple frames to a certain extent. Hence, we choose the fire rate and the membrane potential as the features.

%, which is harder as there will be more data points sharing the same shape to disturb the 

For a data point $x \in R^{T \times 2 \times H \times W}$ ($T$, 2, $H$, and $W$ are the time steps, positive and negative channels, height, and width of the input), the target SNN $M$ outputs the spiking times among $T$ steps as $M(x) \in R^{T \times n}$ ($n$ is the number of classification categories), each of which represents whether a spiking neuron evokes a spike at current time spot. Therefore, the fire rates of $M$ on $x$ is $Fr(x)=\frac{\sum_{t=1}^T M(x)_{t}}{T} \in R^{n}$ and $M(x)_{t} \in R^{n}$ is spiking times at time spot $t$. For the last layer composed of spiking neurons (the number of neurons is $m$), the membrane potential of spiking neurons for $x$ is $Mp(x) \in R^{T \times m}$. Therefore, the average membrane potential among $T$ time steps is $AMp(x)=\frac{\sum_{t=1}^T Mp_{t}(x)}{T}$ and $Mp_{t}(x) \in R^{m}$ is the membrane potential at time step $t$. We utilize fire rates and average membrane potential of the target SNN $M$ on a data point as the signal to predict its membership.

%\textcolor{red}{Usually, $n$ equals $m$ without the inclusion of strategy like voting layer~\cite{Fan_2021_Learnable_Membrane_Time_Constant}, which average the outputs of 10 neurons as the output for a category.} (the last layer with $m$ ReLU neurons in ANNs before conversion)
% Usually, $n$ equals $m$ without including a strategy like voting layer~\cite{Fan_2021_Learnable_Membrane_Time_Constant}, which averages the outputs of 10 neurons as the output for a category.

\subsection{Threat Model}
\label{mem_threat_model}

Following the common practice of MIA~\cite{shokri_membership_2017}, the adversary holds a shadow dataset from the same distribution as a target dataset used for training and testing the target SNN $M$. There are no overlapping data points between the shadow and target datasets. The adversary knows $M$'s hyperparameters and model structure to train a shadow SNN $M_{s}$ for mimicking $M$, following previous works~\cite{shokri_membership_2017,carlini_membership_2021}. Controlling the training and test data of $M_{s}$, the adversary can extract attack features from the training and test data of $M_{s}$ by feeding data into $M_s$ and label features as members (from training data) and non-members (from test data) separately. With the attack features and corresponding labels, the adversary trains a classifier or determines a threshold for distinguishing members and non-members. With the classifier (attack model) or threshold, the adversary evaluates the performance of MIA on the target SNN $M$. 
For attack features, we assume the adversary knows the fire rates $Fr(x)$ and the average membrane potential $AMp(x)$ of a data point $x$ to evaluate the membership privacy of $M$ under various MIAs. 
Finally, the adversary knows the loss, prediction label, and ground truth $y$ of $x$, following the assumptions in previous MIAs on ANNs~\cite{shokri_membership_2017,yeom_privacy_2018}.

\subsection{Methodology}
\label{evaluation_methodology}

We follow strategies of previous MIAs (with confidence scores) on ANNs to leverage fire rates to implement MIAs on SNNs. The reason is that the fire rate is a similar indicator to the confidence score to show the confidence of the prediction. The training of backpropagation-based SNNs aligns the spiking neurons' output with the one-hot encoding representation of the target category with MSE. The alignment aims for SNN to only output a spike in the neuron belonging to the target category at each time step. Ideally, the fire rate of the neuron belonging to the target category is one, and other neurons have a fire rate close to zero. Hence, the fire rate reflects the SNN's confidence in the prediction of the data point. In the previous work~\cite{shokri_membership_2017}, the ANN's prediction confidence gap on the training and test data makes MIAs feasible. Therefore, we follow previous strategies when handling confidence scores to use fire rates. We select five representative methods in the field of MIA, including the first MIA on machine learning~\cite{shokri_membership_2017}, two methods relaxing assumptions of MIA~\cite{salem_ml-leaks_2018}, the performance improvement with modified entropy (Mentr)~\cite{song_systematic_2021}, and performance improvement with logit-scaled confidence~\cite{carlini_membership_2021}.

Apart from the five mentioned methods, we also choose two methods from the work of Yeom et al.~\cite{yeom_privacy_2018}. One is the MIA based on loss, and the other is based on the prediction correctness. The reason for selection is that the loss of the training data tends to be lower than the loss of the test data after training, and the MIA based on prediction correctness provides a baseline method to directly infer the data points correctly predicted by the target model as the training data. To compare the vulnerability of MIAs on ANNs and SNNs, we apply those seven methods. For the average membrane potential, we take it as the feature of the attack model to directly infer the membership following the basic strategy in~\cite{shokri_membership_2017}. For MIAs on ANNs, we also evaluate hinge loss based on logits (the output before the softmax layer in the ANN) due to its more straightforward computation than the logit-scaled confidence and competitive performance~\cite{carlini_membership_2021}.

% \textcolor{red}{The fire rates in SNNs effectively replace confidence scores in ANNs. Considering the training of backpropagation-based SNNs, we get spiking neurons' output (spike or not) in the last classification layer. The number of spiking neurons in the last classification layer equals the number of the classification categories. To improve the performance of SNNs on classification, Mean Square Error (MSE) aligns the spiking neurons' output to the one-hot encoding representation of the target category. The alignment would like the SNN to only output a spike in the neuron belonging to the target category at each time step. Ideally, the fire rate of the neuron belonging to the target category is one, and other neurons have a fire rate close to zero. Therefore, the fire rate is a similar indicator to the confidence score to show the confidence of the prediction.}

In summary, we evaluate MIAs on SNNs with eight methods based on fire rates and membrane potential. For comparison, we implement eight MIAs on ANNs based on (1) confidence scores~\cite{shokri_membership_2017}, (2) loss~\cite{yeom_privacy_2018}, (3) prediction correctness~\cite{yeom_privacy_2018}, (4) top-3 confidence scores~\cite{salem_ml-leaks_2018}, (5) maximum confidence score~\cite{salem_ml-leaks_2018}, (6) logit-scaled confidence~\cite{carlini_membership_2021}, (7) hinge loss~\cite{carlini_membership_2021}, and (8) Mentr with confidence scores~\cite{song_systematic_2021}. For MIAs against SNNs, we discuss the details of each method as follows.

% To evaluate the membership privacy of SNNs, we utilize eight MIAs based on the adversary's knowledge and the SNNs' outputs. Most of the MIAs are inspired by previous works against ANNs. Hence, we compare MIAs against ANNs with MIAs against SNNs. For MIAs against ANNs, we utilize eight methods\todo{g: why those 8?--response: explain in the previous two paragraphs.}, including (1) confidence scores~\cite{shokri_membership_2017}, (2) loss~\cite{yeom_privacy_2018}, (3) prediction correctness~\cite{yeom_privacy_2018}, (4) top-3 confidence scores~\cite{salem_ml-leaks_2018}, (5) maximum confidence score~\cite{salem_ml-leaks_2018}, (6) logit-scaled confidence~\cite{carlini_membership_2021}, (7) hinge loss~\cite{carlini_membership_2021}, and (8) Mentr with confidence scores~\cite{song_systematic_2021}. For MIAs against SNNs, we \textcolor{red}{illustrate}\todo{you really like to use the word expose. But why? What do you want to say with it?} the details of each method as follows.

\textbf{(1) fire rates.} We take the fire rates $Fr(x)$ of each data point as the feature of the attack classifier to predict the membership of each data point.
%The threshold selection is on the shadow model and its training and test data.
\textbf{(2) loss.} We compare the single loss related to each data point with a threshold to determine its membership. The adversary evaluates MIAs on the shadow model and its training and test data to select the threshold that better distinguishes losses of training and test data of the shadow model.

\textbf{(3) prediction correctness.} If the target SNN $M$ correctly predicts the classification label of $x$, we predict $x$ is a member and non-member otherwise.

\textbf{(4) top-3 fire rates.} Instead of using all the fire rates, we select top-3 fire rates as the feature of the attack classifier, following the strategy in the previous work~\cite{salem_ml-leaks_2018}.

\textbf{(5) maximum fire rate.} We use the maximum fire rate to compare with a threshold. If the maximum fire rate exceeds the threshold, we predict $x$ as member and non-member otherwise.

\textbf{(6) logit-scaled fire rate.} We compute the logit-scaled fire rate as Eq.~\eqref{logit-scaled_fire_rate}, where $Fr(x)_y$ is the fire rate of the spiking neuron of the ground truth. We replace the confidence score of logit-scaled confidence~\cite{carlini_membership_2021} with the fire rate to obtain this metric. Similarly, we compare the logit-scaled fire rate with a threshold for the membership prediction.

\begin{equation}
    \label{logit-scaled_fire_rate}
    \textit{$\log(Fr(x)_{y})-\log \sum_{y^{'} \neq y} Fr(x)_{y^{'}}$.}
\end{equation}

\textbf{(7) Mentr with fire rates.} Following the Mentr with confidence scores in the previous work~\cite{song_systematic_2021}, we calculate the Mentr with fire rates as Eq.~\eqref{Mentr_with_fire_rates}. We compare this metric with a threshold while deciding on a membership of $x$.

\begin{equation}
    \label{Mentr_with_fire_rates}
    \textit{$-(1-Fr(x)_{y})\log(Fr(x)_{y})-\sum_{y^{'} \neq y} Fr(x)_{y^{'}} \log(1-Fr(x)_{y^{'}})$.}
\end{equation}

\textbf{(8) average membrane potential.} Apart from fire rates, we utilize the average membrane potential as the feature of the attack classifier to predict membership. Similarly, we train the attack classifier with features extracted from the shadow model on its data.

%% file: sections/experiments.tex
\section{Experimental Settings}
\label{sec:experiments}

We discuss the datasets used in our experiments in Section~\ref{dataset_explain}. Then, we discuss the ANN and SNN models in Section~\ref{models_explain}. In Section~\ref{settings_explain}, we detail the settings of training models, including target, shadow, and attack models. Besides, we clarify the evaluation metric.

\subsection{Datasets}
\label{dataset_explain}

We select three neuromorphic datasets, N-MNIST~\cite{orchard2015converting}, CIFAR10-DVS~\cite{li2017cifar10}, and N-Caltech101~\cite{orchard2015converting}, to explore MIAs against SNNs. For comparison, we choose corresponding static versions (i.e., MNIST, CIFAR10, and Caltech101) of those three datasets to train ANNs. Table~\ref{num_data_points} shows the shape of a batch of data points and the number of data points in each dataset. Among the numbers representing the shape, the $T$, $B$, and the last two numbers separately indicate the time steps, batch size, and height and width of data points.

\begin{table}[ht]%[ht]
\centering
\caption{Statistical information of each dataset.}
\label{num_data_points}
\begin{adjustbox}{max width=0.48\textwidth}
\begin{tabular}{ccc}
\toprule
% \multirow{2}{*}{Dataset}  & \multirow{2}{*}{\begin{tabular}[c]{@{}c@{}}Feature \\ Length\end{tabular}}  & \multirow{2}{*}{\begin{tabular}[c]{@{}c@{}}Total Subject \\ Number\end{tabular}} & \multirow{2}{*}{\begin{tabular}[c]{@{}c@{}}Per-Subject \\ Data Points Range\end{tabular}}  & \multirow{2}{*}{\begin{tabular}[c]{@{}c@{}}Subject \\ Number\end{tabular}}\\
Dataset  & Shape of a batch of data points & Number of data points\\
% & & & &\\
\cmidrule{1-3}
N-MNIST& $T\times B\times 2\times 34\times 34$ & 70,000\\
CIFAR10-DVS& $T\times B\times 2\times 128\times 128$ & 10,000 \\
N-Caltech101& $T\times B\times 2\times 180\times 240$ & 9,146 \\
MNIST& $B\times 3\times 28\times 28$ & 70,000  \\
CIFAR-10& $B\times 3\times 32\times 32$ & 60,000 \\
Caltech101& $B\times 3\times 180\times 240$ (resize) & 9,146 \\
\bottomrule
\end{tabular}
\end{adjustbox}
\end{table}

We load neuromorphic datasets from SpikingJelly~\cite{fang2023spikingjelly} and static datasets from Pytorch. For dividing the whole dataset into two sub-datasets for shadow and target models, we select 50\% data points as the target model's sub-dataset (training and test) and the remaining 50\% as the shadow model's sub-dataset (training and test). 
For N-Caltech101 and Caltech101, the platform does not already divide the training and test data. Thus, we sample 90\% data points of each sub-dataset as the training data and the left 10\% as the test data to ensure the training data of models. For the four remaining datasets, we will follow the data division of the platform to sample training and test data. 
Besides, we keep the same number of member and non-member data for training the attack model and evaluating the performance of MIAs on the target model. 

Apart from training SNNs with neuromorphic datasets and ANNs with static datasets, we train ANNs with neuromorphic datasets and SNNs with static datasets. In the work of Deng et al.~\cite{Deng_2020_Rethink}, the authors trained ANNs with neuromorphic datasets and SNNs with static datasets to empirically compare their performance on the visual recognition task. Inspired by their work, we investigate the membership privacy of ANNs with neuromorphic datasets and SNNs with static datasets, considering the feasibility of training.
For a batch of data points ($R^{T \times B \times 2 \times H \times W}$) in the neuromorphic dataset, we follow the previous work~\cite{Deng_2020_Rethink} to normalize the accumulated spike times to the image pixel range (0 to 1 after normalization) by dividing by the maximum spiking time of each location in this batch of data points and labeling each frame of the data point as the same label. This means that we keep positive and negative channels in the neuromorphic dataset rather than inserting a full-zero channel to construct an RGB format image like the ``play\_frame'' function of SpikingJelly. This is done to reduce the memory consumption of training and keep the attack performance comparison of the SNN and ANN both trained with the neuromorphic dataset fair. For a data point from the static dataset, we repeat the image for time-step times and utilize the first few layers of the SNN to transfer the static image into spike events~\cite{Fan_2021_Learnable_Membrane_Time_Constant} rather than transferring with a Poisson encoder, which might incur variability in the firing of the network and impair its performance~\cite{rueckauer2017conversion}.

% For static datasets, we can repeat the image for time-step times and utilize a spiking encoder network as~\cite{Fan_2021_Learnable_Membrane_Time_Constant} to transfer the static image into spike trains rather than transferring with a Poisson encoder, which might incur variability into the firing of the network and impair its performance~\cite{rueckauer2017conversion}.

% How to fairly compare the SNN and DNN is a vital problem in this work.
% (1) One static image is transferred into spike trains and put into SNN; change the spiking neuron to an activation neuron, keep others unchanged, and remove the time steps.
% (2) Event-driven neuromorphic data (formate as [T, B, C, H, W], time steps, batch size, channel, height, and width): each time step, the input is an image summarized with multiple events, we label the frame in each time step with the same label and use the DNN to train with them. It is a way to transfer neuromorphic data for DNN.
% (3) To make a fair comparison, we do not apply augmentations apart from basic transforms (resizing and scaling the pixel value range) to train the ANNs and SNNs. Training the ANN with complex augmentations will enhance the generalization and reduce the performance of MIAs; we avoid complex augmentations for comparison and explore them as a defense strategy.

\subsection{Models}
\label{models_explain}

We select three model structures to train SNNs, including the model defined in the work of Fang et al.~\cite{Fan_2021_Learnable_Membrane_Time_Constant} (we denote it CNN for convenience), VGG11~\cite{simonyan2014very}, and ResNet18~\cite{he2016deep}. The number of convolutional, downsampling, and fully connected layers (i.e., $N_{conv}$, $N_{down}$, and $N_{fc}$) in CNN from the work of Fang et al.~\cite{Fan_2021_Learnable_Membrane_Time_Constant} is given in Table~\ref{num_layer_info}. Each downsampling layer comprises $N_{conv}$ convolutional layers and a max-pooling layer. $N_{down}$ downsampling layers and $N_{fc}$ fully connected layers make the final model. For N-Caltech101, we modified the model structure for DVS128 Gesture from~\cite{Fan_2021_Learnable_Membrane_Time_Constant} to fit the input size of N-Caltech101. 
For ANNs with structures of VGG11 and ResNet18, we replace the neuron formulated as a ReLU activation function with spiking neurons to construct the corresponding SNNs with structures of VGG11 and ResNet18. We keep the weights and connections between neurons. Considering that there are two input channels for the neuromorphic data and that the original input of VGG11 and ResNet18 is expected to have three channels, we add a convolutional layer to increase the number of channels (upsampling).

% For VGG11 and ResNet18, we replaced the artificial neuron with a ReLU activation function with the Spiking neuron and keep other parts to train SNNs.\todo{???} 

\begin{table}[ht]%[ht]
\centering
\caption{Number of different layers in a CNN.}
\label{num_layer_info}
\begin{adjustbox}{max width=0.3\textwidth}
\begin{tabular}{cccc}
\toprule
% \multirow{2}{*}{Dataset}  & \multirow{2}{*}{\begin{tabular}[c]{@{}c@{}}Feature \\ Length\end{tabular}}  & \multirow{2}{*}{\begin{tabular}[c]{@{}c@{}}Total Subject \\ Number\end{tabular}} & \multirow{2}{*}{\begin{tabular}[c]{@{}c@{}}Per-Subject \\ Data Points Range\end{tabular}}  & \multirow{2}{*}{\begin{tabular}[c]{@{}c@{}}Subject \\ Number\end{tabular}}\\
Dataset  & $N_{conv}$  & $N_{down}$ & $N_{fc}$\\
% & & & &\\
\cmidrule{1-4}
N-MNIST&  1  & 2 & 2 \\
CIFAR10-DVS& 1 & 4 & 2\\
N-Caltech101& 1 & 5 & 2\\
\bottomrule
\end{tabular}
\end{adjustbox}
\end{table}

For the CNN originally defined for neuromorphic datasets in the work of Fang et al.~\cite{Fan_2021_Learnable_Membrane_Time_Constant}, we modify the spiking neurons to artificial neurons with a ReLU activation function and change the input channel of the first convolutional layer to three as the number of channels in the static data is three. It is the opposite process compared to modifying the ANN to the SNN. 
Note that instead of directly utilizing the previous CNN (like AlexNet~\cite{krizhevsky2012imagenet}) to train on static datasets, we modify the CNN originally defined for neuromorphic datasets (SNN) to the CNN used for static datasets. This modification allows a fair comparison between SNNs trained with neuromorphic datasets and ANNs trained with static datasets.
For conversion from ANNs to SNNs, we discussed the conversion operation in Section~\ref{background_SNN}.
% the CNN originally defined for neuromorphic
% For three static datasets, we replaced the Spiking neurons in the models for the neuromorphic datasets with artificial neurons with a ReLU activation function and kept other parts, which leaves an almost close model structure for a fair comparison between SNNs trained with neuromorphic datasets and ANNs trained with static datasets.\todo{these sentences make no sense} 

For the classifier-based MIA, the attack model is a multilayer perceptron (MLP) with 2 hidden layers, each with 64 neurons, following the previous work~\cite{label_only_MIA_Mauro}. The single-value output of the MLP represents the probability of being predicted as a member.

\subsection{Settings and Evaluation Metric}
\label{settings_explain}

% Default training settings of target models, including the epoch number, learning rate, optimizer, time steps, 
For ANNs and SNNs trained with neuromorphic and static datasets, per default, we utilize the identical learning rate of 0.001, Adam optimizer, and batch size range from 2 to 16 due to the input size. For ANNs, the loss function is cross-entropy loss, while the loss function is Mean Squared Error (MSE) for SNNs. The number of epochs is 30, 50, and 60 for MNIST (N-MNIST), CIFAR-10 (CIFAR10-DVS), and Caltech101 (N-Caltech101), respectively. The default number of time steps for ANNs is 16. For the attack model, the optimization algorithm is Adam, with a learning rate of 0.001. The number of training epochs is 300, and the batch size is 32. We use the Binary Cross Entropy (BCE) to guide the training of the attack model. In the hyperparameter study in Section~\ref{ablation_study}, we explore the impact of the optimizer, learning rate, and time steps on the performance of target models and MIAs, as they are vital settings for training SNNs.

For the evaluation metric of MIAs, we follow the previous works~\cite{shokri_membership_2017,carlini_membership_2021} and use the balanced accuracy as the comparison metric of MIAs. %\textcolor{red}{From our experiments, the balanced accuracy can effectively measure the membership privacy of SNNs and compare the vulnerability with ANNs.}

%% file: sections/results_and_discussion.tex
\section{Results and Discussions}
\label{sec:results}

We compare the performance of MIAs on ANNs and SNNs in Section~\ref{SNN_vs_ANN}. Next, we provide a hyperparameter study in Section~\ref{ablation_study}. Finally, we evaluate the basic augmentation method for static datasets and two augmentation mechanisms for neuromorphic datasets as the defenses against MIAs in SNNs in Section~\ref{defense}.

\subsection{MIAs on SNNs and ANNs}
\label{SNN_vs_ANN}

We first compare the vulnerability of SNNs and ANNs under MIAs. To make the comparison fair, we make the structures of SNNs and ANNs similar except for the neurons. The hyperparameters (time steps, learning rate, batch size, and dataset split) are identical when utilizing the same dataset. We do not leverage data augmentation or other techniques to improve the generalization of models here, as those techniques usually differ between ANNs and SNNs~\cite{Li_NDA_2022}. 
However, we explore how to defend against MIAs with data augmentation mechanisms by improving the generalization of models in Section~\ref{defense}.

\begin{figure}
    % \Description[MIAs on static datasets]
    \centering
    \begin{subfigure}[b]{0.23\textwidth}
        \centering
        \includegraphics[scale=0.46]{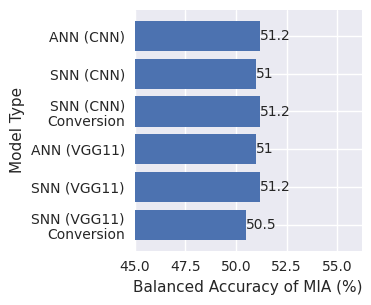}
        \caption{MIAs (MNIST)}
        \label{mnist_ann_snn}
    \end{subfigure}
    \begin{subfigure}[b]{0.23\textwidth}
        \centering
        \includegraphics[scale=0.46]{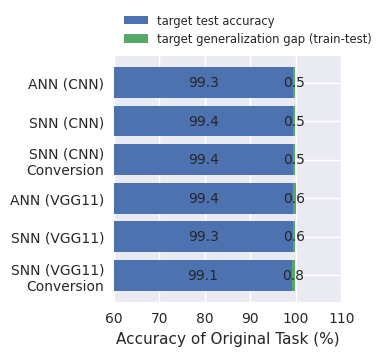}
        \caption{Target Models (MNIST)}
        \label{mnist_target_acc}
    \end{subfigure}
    
    \begin{subfigure}[b]{0.23\textwidth}
        \centering
        \includegraphics[scale=0.45]{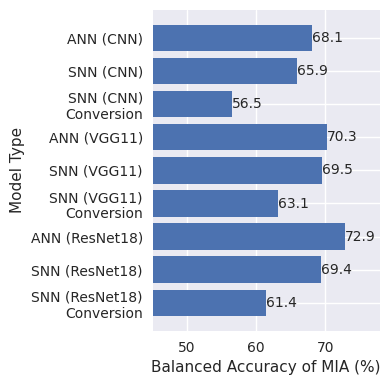}
        \caption{MIAs (CIFAR-10)}
        \label{cifar-10_ann_snn}
    \end{subfigure}
    \begin{subfigure}[b]{0.23\textwidth}
        \centering
        \includegraphics[scale=0.46]{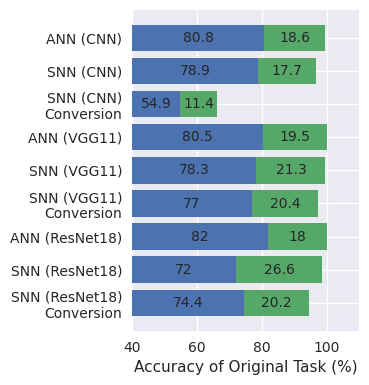}
        \caption{Target Models (CIFAR-10)}
        \label{cifar-10_target_acc}
    \end{subfigure}
    
    \begin{subfigure}[b]{0.23\textwidth}
        \centering
        \includegraphics[scale=0.44]{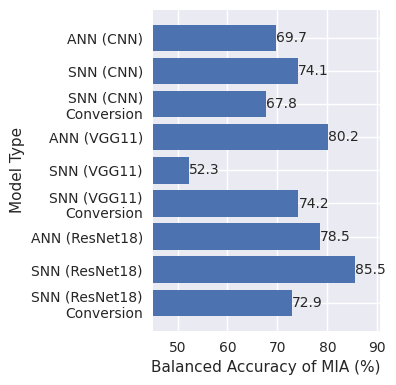}
        \caption{MIAs (Caltech101)}
        \label{caltech101_ann_snn}
    \end{subfigure}
    \begin{subfigure}[b]{0.23\textwidth}
        \centering
        \includegraphics[scale=0.46]{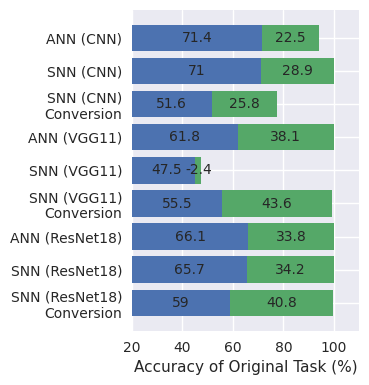}
        \caption{Target Models (Caltech101)}
        \label{caltech_target_acc}
    \end{subfigure}
   
    \caption{The highest accuracy of eight MIAs, target test accuracy, and target generalization gap of three static datasets with various model types.}
    \label{static_highest_acc_of_eight_MIAs}
\end{figure}

\begin{figure}
    % \Description[MIAs on neuromorphic datasets]
    \centering
    \begin{subfigure}[b]{0.23\textwidth}
        \centering
        \includegraphics[scale=0.46]{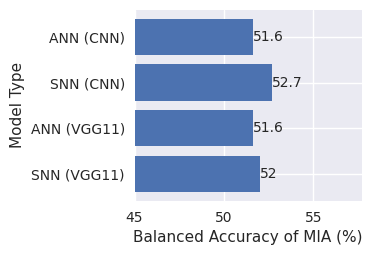}
        \caption{MIAs (N-MNIST)}
        \label{nmnist_ann_snn}
    \end{subfigure}
    \begin{subfigure}[b]{0.23\textwidth}
        \centering
        \includegraphics[scale=0.46]{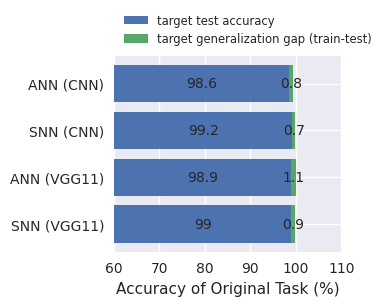}
        \caption{Targets (N-MNIST)}
        \label{nmnist_target_acc}
    \end{subfigure}
    
    \begin{subfigure}[b]{0.23\textwidth}
        \centering
        \includegraphics[scale=0.435]{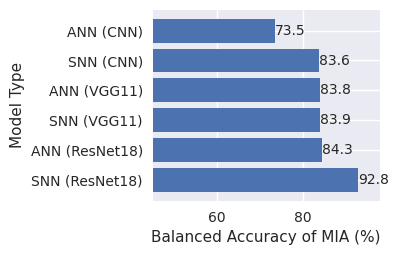}
        \caption{MIAs (CIFAR10-DVS)}
        \label{cifar-10-dvs_ann_snn}
    \end{subfigure}
    \begin{subfigure}[b]{0.23\textwidth}
        \centering
        \includegraphics[scale=0.46]{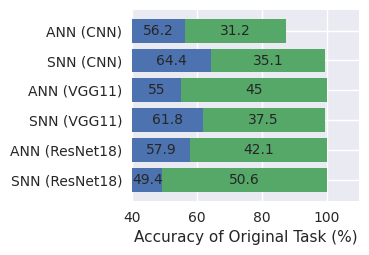}
        \caption{Targets (CIFAR10-DVS)}
        \label{cifar-10-dvs_target_acc}
    \end{subfigure}
    
    \begin{subfigure}[b]{0.23\textwidth}
        \centering
        \includegraphics[scale=0.445]{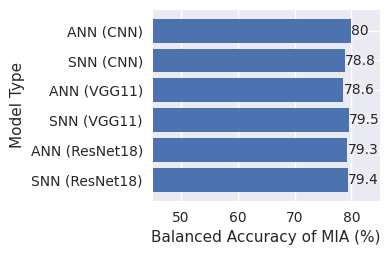}
        \caption{MIAs (N-Caltech101)}
        \label{ncaltech101_ann_snn}
    \end{subfigure}
    \begin{subfigure}[b]{0.23\textwidth}
        \centering
        \includegraphics[scale=0.46]{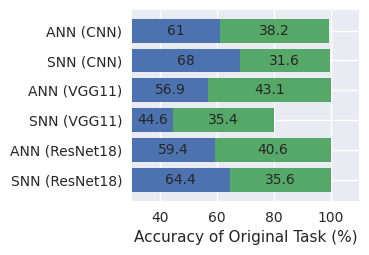}
        \caption{Targets (N-Caltech101)}
        \label{ncaltech_target_acc}
    \end{subfigure}
   
    \caption{The highest accuracy of eight MIAs, target test accuracy, and target generalization gap of three neuromorphic datasets with various model types.}
    \label{neuromorphic_highest_acc_of_eight_MIAs}
\end{figure}

Figures~\ref{static_highest_acc_of_eight_MIAs} and~\ref{neuromorphic_highest_acc_of_eight_MIAs} show the highest accuracy of eight MIAs and the performance of the original classification task under target models for three static and three neuromorphic datasets. In the left subfigure, the x-axis represents the highest balanced accuracy among eight MIAs, and the y-axis indicates the model type, among which conversion means the ANN is converted to SNN. In the right subfigure, the x-axis represents the accuracy of the original classification task, and the y-axis represents the model type. We show the accuracy of the original task via the target test accuracy and target generalization gap, which is target training accuracy minus target test accuracy. From previous works~\cite{yeom_privacy_2018,shokri_membership_2017}, a higher generalization gap usually leads to a larger privacy leakage, i.e., a higher attack performance.

% We speculate the property of the dataset makes the training of models and their vulnerability different. We leave it as future work to deeply investigate what kind of dataset properties contribute to the phenomenon we observe.
% \textcolor{red}{which indicates the perporties of the static dataset impact the vulnerability comparison}

\textbf{Comparing the highest attack accuracy of MIAs among ANNs and SNNs under the same static dataset, we find vulnerability is related to the type of static dataset}.
As the training data of the target model only occupy half of the original training data, the original classification accuracy of the target model is relatively lower than the accuracy reported in the previous work~\cite{Fan_2021_Learnable_Membrane_Time_Constant}. For MNIST, the highest attack accuracy of ANNs is similar to that of SNNs (around 50\%) as the training and test accuracy of the target model is all above 99\%, indicating almost no generalization gap. For CIFAR-10, the highest attack accuracy of ANNs is larger than that of SNNs, even though SNNs have a larger generalization gap than ANNs. For example, the SNN with a structure of ResNet18 has a generalization gap of 26.6\% (larger than 18\% with the ANN) and the highest attack accuracy of 69.4\% (smaller than 72.9\% with the ANN). We also observe this trend with CNN and VGG11 trained with CIFAR-10. Besides, the larger generalization gap does not always represent a higher attack performance under MIAs. The ANN with a lower generalization gap could have a higher attack accuracy than not only the SNN with a higher generalization gap but also the ANN with a higher generalization gap. 
This is an extension of previous findings~\cite{shokri_membership_2017,carlini_membership_2021,conti2022vulnerability} since they observed that ANN with a high generalization gap could have a low attack performance under MIAs. 

For Caltech101, MIAs obtain a higher attack accuracy on SNNs than ANNs. For example, the SNN with a structure of CNN has the highest attack accuracy of 74.1\%, higher than 69.7\% of the ANN. Similarly, MIAs have higher attack performance on SNN than ANN when the structure is ResNet18. While training different models with the same static dataset, the vulnerability comparison of SNNs and ANNs is consistent, which means the model type is not the factor that leads to the difference. % in vulnerability comparison with various static datasets. 
Therefore, the dataset itself is the reason, corresponding with the previous work~\cite{conti2022vulnerability} that finds the dataset impacts the performance of MIAs. We notice the SNN with a structure of VGG11 has a low original classification accuracy (around 48\%) and a negative generalization gap, which leads to the low performance of MIAs (52.3\%). In summary, the vulnerability comparison between SNNs and ANNs depends on the static datasets used for training ANNs and SNNs. %\textcolor{red}{The ANN with a low generalization gap could be more vulnerable than the SNN or ANN with a high generalization gap, an extended finding to previous works.} 

%Thus, for VGG11 and ResNet18, conversion from ANNs to SNNs maximally reduces 11.5\% on the performance of MIAs with a drop of 7.6\% on the original classification task. The reduction in the accuracy of MIAs could maximally be twice the decrease in the accuracy of the original classification task.
% Besides, the reduction (7.2\% for VGG11 with CIFAR-10) of the highest accuracy of MIAs could be maximally twice the decrease (3.5\% for VGG11 with CIFAR-10) of the original classification accuracy.

\textbf{Conversion from ANNs to SNNs reduces the performance of MIAs against converted SNNs with the accuracy drop on the original classification task.} For ANNs with a structure of CNN, the conversion to SNNs has a reduction of about 20\% on the original classification accuracy on CIFAR-10 and Caltech101. The highest accuracy of MIAs drops from 68.1\% to 56.5\% (11.6\%) for CIFAR-10 and from 69.7\% to 67.8\% (1.9\%) for Caltech101. For VGG11 and ResNet18, converting from ANNs to SNNs brings an original classification accuracy drop within 8\% (CIFAR-10 and Caltech101). However, the reduction of the highest attack accuracy could be larger than 10\%. For example, the conversion from the ANN (ResNet18) with the highest attack accuracy of 72.9\% to the SNN obtains a smaller highest attack accuracy of 61.4\% (a reduction of 11.5\%) with the original classification accuracy decreased from 82\% to 74.4\% (a reduction of 7.6\%), even though the generalization gap increases from 18\% to 20.2\%.
There are two reasons for the drop in MIAs' performance. The first reason is the decrease in performance of the original classification task and the generalization gap. The second is that converted SNN utilizes different neurons to connect weights and forward the input compared to the original ANN, which means a different way of weight utilization compared to the training of weights. As the weights are related to the training data and membership information, converted SNN loses part of the information due to the change of weight utilization, reducing the performance of MIAs and the original classification task. 
Since conversion from ANNs to SNNs could bring lower membership privacy leakage and a relatively low drop in the original task, we recommend the conversion to protect data points' privacy.
%acceptable

\textbf{When using neuromorphic datasets, SNNs are more vulnerable than ANNs due to the larger generalization gap.} For example, the SNN with a structure of ResNet18 trained with CIFAR10-DVS has a generalization gap of 50.6\%, which is higher than 42.1\% with the ANN. Under this condition, the highest accuracy of MIAs for the SNN is 92.8\%, which is higher than 84.3\% with the ANN. With N-Caltech101, the highest accuracy of MIAs is similar for ANNs and SNNs. However, the generalization gaps of SNNs are smaller than those of ANNs. For example, the SNN with a structure of ResNet18 trained with N-Caltech101 has a generalization gap of 35.6\%, which is smaller than 40.6\% with the ANN. Nevertheless, the SNN and ANN have a similar highest attack accuracy of 79\%. The SNN obtains a similar highest attack accuracy with a smaller generalization gap than the ANN, while the higher generalization gap usually leads to a higher attack accuracy. Hence, MIAs have a higher attack accuracy on SNNs than on ANNs due to a larger generalization gap or a similar attack accuracy on SNNs and ANNs if the generalization gap on SNNs is smaller than on ANNs. From the previous analysis, we conclude that the vulnerability of SNNs is higher than that of ANNs while being trained with neuromorphic datasets.
%This indicates that SNNs are more vulnerable than ANNs while trained with neuromorphic datasets.

\begin{figure}[ht!]
    % \Description[generalization gap]
    \centering
    \includegraphics[scale=0.6]{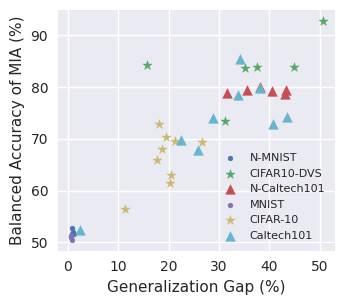}
    \caption{The relationship between the generalization gap and the highest accuracy among eight MIAs.}
    \label{generalization_gap_MIA}
\end{figure}

Figure~\ref{generalization_gap_MIA} presents the relationship between the generalization gap and the highest accuracy among eight MIAs. As we can observe, the higher generalization gap usually leads to higher vulnerability under the MIAs, which corresponds to the conclusion from previous works~\cite{shokri_membership_2017,yeom_privacy_2018}. \textbf{This indicates the rule that a higher generalization gap of the target model usually leads to a higher vulnerability toward MIAs is still applicable for SNNs.}

We show the attack accuracy of eight MIAs in Appendices~\ref{MIAs_against_SNNs} and~\ref{MIAs_against_ANNs}.
\textbf{From these two tables, we observe that loss, logit-scaled confidence (fire rate), and Mentr with confidence scores (fire rates) are the three metrics with higher accuracy among the metrics explored.} For example, both MIAs based on those three metrics obtain the highest accuracy of 78.6\% against the ANN (VGG11) trained with N-Caltech101.

To investigate the modification of training loss and attack accuracy of MIAs along the training epochs, we apply MIAs against the trained model obtained after each epoch. Figure~\ref{modification_along_epochs} in Appendix~\ref{accuracy_along_epoch} shows how the loss, accuracy, and attack accuracy change during the training epochs of the target and shadow models. We consider the ANN and SNN based on CNN with CIFAR-10 and CIFAR10-DVS as examples. The figure shows that the SNN has a stable test loss after reaching the minimum test loss. However, the test loss of the ANN will increase after arriving at the minimum test loss. The generalization gap increases during the training of ANNs and SNNs. Finally, the test accuracy becomes stable. As the optimization, the training accuracy still slowly increases with more epochs. \textbf{The accuracy of MIAs increases along the training epochs due to the increase of the generalization gap.}

% \subsection{Improved MIA}
% Considering the data format and pipeline of the SNN, the time step $T$ during the model inference is not strictly equal to the one used in the training. This means that we can segment the events into different numbers of frames, each of which is the input data of one step in the model training and inference. Different numbers of steps lead to the variance of the spiking times, which might be the signal to distinguish the training and test data of the target SNN. This is a unique way of implementing the MIA against SNN.

\subsection{Hyperparameter Study}
\label{ablation_study}

To explore the factors that impact the performance of MIAs on SNNs, we conduct a hyperparameter study to vary the spiking neuron type, surrogate function, optimizer, learning rate, and time steps during the training of SNNs.

\subsubsection{Varying Spiking Neuron Type and Surrogate Function}

For the backpropagation-based SNNs, we not only train the SNN with the LIF neuron type and the surrogate function of ATan to get the results in Table~\ref{MIAs_SNNs} but also explore other neuron types and surrogate functions. 
We select neuromorphic datasets for training SNNs as examples to analyze the impact of spiking neuron type and surrogate function. We utilize three spiking neuron types (LIF, EIF~\cite{fourcaud2003spike}, and Izhikevich~\cite{izhikevich2003simple}) and two surrogate functions (ATan and PiecewiseLeakyReLU~\cite{Wu_2018_STBP}). The reason for selecting those three spiking neuron types and two surrogate functions is that they are common in the research of SNNs~\cite{fourcaud2003spike,izhikevich2003simple,Wu_2018_STBP,Neftci_Surrogate_2019,tavanaei2019deep}.

Table~\ref{vary_neuron_surrogate_nmnist} provides the accuracy of the target model and the MIA with the highest accuracy under each setting. For N-MNIST, the SNN with PiecewiseLeakyReLU function and Izhikevich neuron is the most vulnerable one with the highest attack accuracy of 55.5\%. For CIFAR10-DVS and N-Caltech101, SNNs with the ATan function and the LIF neuron have the highest attack accuracy (83.6\% in CIFAR10-DVS and 77.6\% in N-Caltech101) compared to other functions and neurons. This is mainly due to the larger generalization gap if the SNN is defined with the ATan function and the LIF neuron, even though the test accuracy is higher than that of other spiking neurons and surrogate functions.

\begin{table}[ht]
\centering
\caption{The attack accuracy under different spiking neuron types and surrogate functions while training SNNs (CNN).}
\label{vary_neuron_surrogate_nmnist}
\begin{adjustbox}{max width=0.48\textwidth}
\begin{tabular}{ccccccc}
\hline
\addlinespace 
\multirow{3}{*}{\begin{tabular}[c]{@{}c@{}}Dataset\end{tabular}} & \multirow{3}{*}{\begin{tabular}[c]{@{}c@{}}Surrogate \\Function\end{tabular}} & \multirow{3}{*}{\begin{tabular}[c]{@{}c@{}}Spiking \\ Neuron\end{tabular}} & \multirow{3}{*}{\begin{tabular}[c]{@{}c@{}}Target\\ Test\\Acc\end{tabular}} & \multirow{3}{*}{\begin{tabular}[c]{@{}c@{}}Target \\Train\\Acc\end{tabular}} &\multicolumn{2}{c}{MIA with Highest Acc}\\ 
% \cline{3-11} 
\cmidrule(lr){6-7} 
\addlinespace 
 & & & & & MIA & Acc \\ 
%\multirow{2}{*}{fire rates} & \multirow{2}{*}{loss} & \multirow{2}{*}{\begin{tabular}[c]{@{}c@{}}prediction \\correctness\end{tabular}} & \multirow{2}{*}{\begin{tabular}[c]{@{}c@{}}top-3 \\fire rates\end{tabular}} & \multirow{2}{*}{\begin{tabular}[c]{@{}c@{}}maximum \\fire rate\end{tabular}} & \multirow{2}{*}{\begin{tabular}[c]{@{}c@{}}logit-scaled \\fire rate\end{tabular}} & \multirow{2}{*}{\begin{tabular}[c]{@{}c@{}}Mentr with \\fire rates\end{tabular}} & \multirow{2}{*}{\begin{tabular}[c]{@{}c@{}}avg membrane\\ potential\end{tabular}}\\ 
% fire rates   loss   prediction \\correctness  top-3 \\fire rates  maximum \\fire rate  logit-scaled \\fire rate  Mentr with \\fire rates avg membrane\\ potential
\addlinespace
\hline
\addlinespace 
\multirow{8}{*}{N-MNIST} &ATan & LIF & 99.2\% & 99.9\% & fire rates & 51.4\%\\
% & TrainInversion & 336   & 48   & 336  & 48  & 336  & 48  & 336  & 48  &    \\
\addlinespace 
& ATan & EIF & 99.2\%& 99.7\%& \begin{tabular}[c]{@{}c@{}}maximum \\fire rate\end{tabular} & 51.6\%\\
% & TrainInversion & 336   & 48   & 336  & 48  & 336  & 48  & 336  & 48  &    \\
\addlinespace 
& ATan & Izhikevich & 98.8\% & 99.6\%& loss & 50.9\%\\
% & TrainInversion & 336   & 48   & 336  & 48  & 336  & 48  & 336  & 48  &    \\
\addlinespace 
& PiecewiseLeakyReLU & LIF & 99.2\% & 99.9\% & loss & 51.7\%\\
% & TrainInversion & 336   & 48   & 336  & 48  & 336  & 48  & 336  & 48  &    \\
\addlinespace 
& PiecewiseLeakyReLU & EIF & 99.1\%& 99.8\%&loss& 51.7\%\\
% & TrainInversion & 336   & 48   & 336  & 48  & 336  & 48  & 336  & 48  &    \\
\addlinespace 
& PiecewiseLeakyReLU & Izhikevich & 98.8\% & 99.3\% & \begin{tabular}[c]{@{}c@{}}Mentr with \\fire rates\end{tabular} &55.5\%\\
\hline

\addlinespace 
\multirow{9}{*}{CIFAR10-DVS} & ATan & LIF & 66.4\% & 99.4\% & loss & 83.6\%\\
% & TrainInversion & 336   & 48   & 336  & 48  & 336  & 48  & 336  & 48  &    \\
\addlinespace 
& ATan & EIF & 56.0\%& 82.0\%& \begin{tabular}[c]{@{}c@{}}logit-scaled \\fire rate\end{tabular}& 69.2\%\\
% & TrainInversion & 336   & 48   & 336  & 48  & 336  & 48  & 336  & 48  &    \\
\addlinespace 
& ATan & Izhikevich & 32.2\% & 30.8\%& \begin{tabular}[c]{@{}c@{}}logit-scaled \\fire rate\end{tabular}& 53.3\%\\
% & TrainInversion & 336   & 48   & 336  & 48  & 336  & 48  & 336  & 48  &    \\
\addlinespace 
& PiecewiseLeakyReLU & LIF & 62.6\% & 97.1\% & loss & 78.7\%\\
% & TrainInversion & 336   & 48   & 336  & 48  & 336  & 48  & 336  & 48  &    \\
\addlinespace 
& PiecewiseLeakyReLU & EIF & 56.4\%& 68.5\%& loss & 61.3\%\\
% & TrainInversion & 336   & 48   & 336  & 48  & 336  & 48  & 336  & 48  &    \\
\addlinespace 
& PiecewiseLeakyReLU & Izhikevich & 33.6\% & 29.6\% & \begin{tabular}[c]{@{}c@{}}Mentr with \\fire rates\end{tabular}& 55.5\%\\
\hline

\addlinespace 
\multirow{8}{*}{N-Caltech101} & ATan & LIF & 69.9\% & 99.8\% & loss & 77.6\%\\
% & TrainInversion & 336   & 48   & 336  & 48  & 336  & 48  & 336  & 48  &    \\
\addlinespace 
& ATan & EIF & 58.2\%& 86.5\%& loss & 69.0\%\\
% & TrainInversion & 336   & 48   & 336  & 48  & 336  & 48  & 336  & 48  &    \\
\addlinespace 
& ATan & Izhikevich & 61.8\% & 81.8\%& loss & 64.3\%\\
% & TrainInversion & 336   & 48   & 336  & 48  & 336  & 48  & 336  & 48  &    \\
\addlinespace 
& PiecewiseLeakyReLU & LIF & 41.6\% & 42.9\% & loss & 55.4\%\\
% & TrainInversion & 336   & 48   & 336  & 48  & 336  & 48  & 336  & 48  &    \\
\addlinespace 
& PiecewiseLeakyReLU & EIF & 28.7\%& 27.7\% & fire rates & 53.0\%\\
% & TrainInversion & 336   & 48   & 336  & 48  & 336  & 48  & 336  & 48  &    \\
\addlinespace 
& PiecewiseLeakyReLU & Izhikevich & 48.7\% & 48.8\% & \begin{tabular}[c]{@{}c@{}}prediction \\correctness\end{tabular} & 58.3\%\\

\hline

\end{tabular}
\end{adjustbox}
\end{table}

\subsubsection{Varying Optimizer and Learning Rate}
% As the training and test accuracy of SNNs on N-MNIST is all close to 100\% under the default setting of the optimizer and learning rate, which is already hard to implement MIAs, we do not explore with N-MNIST here.
%In N-Caltech101, a higher generalization gap (35.4\%) of the SNN trained with Adam and a learning rate of 0.001 leads to a higher attack accuracy of 79.5\% than the SNN trained with SGD and a learning rate of 0.1.
As the selection of optimizer and learning rate impacts the generalization of the final trained model, we explore their influence on the final accuracy of SNNs and the performance of MIAs on SNNs. Table~\ref{vary_optimizer_and_learning_rate} provides the accuracy of the target model and the MIA with the highest accuracy while training with different optimizers and learning rates. The table shows two settings unsuitable for training because their final classification accuracy is similar to random guessing, including Adam, with a learning rate of 0.1, and SGD, with a learning rate of 0.001. Indeed, the attack accuracy of MIAs is close to 50\% under those two settings. 
For N-Caltech101, a higher generalization gap (35.4\%) of the SNN trained with Adam and a learning rate of 0.001 leads to a higher attack accuracy of 79.5\% compared to the SNN trained with SGD and a learning rate of 0.1. 
For CIFAR10-DVS, the SNN optimized with Adam and a learning rate of 0.001 has a higher generalization gap (37.5\%) than the SNN trained with the SGD and a learning rate of 0.1 (34.7\%). However, MIAs have a smaller attack accuracy (83.9\%) on the SNN trained with Adam and a learning rate of 0.001. This shows that a higher generalization does not always indicate a higher attack accuracy. From the training and test accuracy of SNNs with different optimizers and learning rates, we also observe that the training of SNNs is sensitive to those hyperparameters. 
A large (0.1) or a small learning rate (0.001) could both lead to the final trained SNN at the random guessing performance. The learning rate suitable for an optimizer might not be suitable for another optimizer, which indicates optimizers and learning rates should be considered together for training SNNs.

\begin{table}[ht]
\centering
\caption{The attack accuracy under different optimizers and learning rates while training SNNs (VGG11).}
\label{vary_optimizer_and_learning_rate}
\begin{adjustbox}{max width=0.46\textwidth}
\begin{tabular}{ccccccc}
\hline
\addlinespace 
\multirow{3}{*}{\begin{tabular}[c]{@{}c@{}}Dataset\end{tabular}} & \multirow{3}{*}{Optimizer} & \multirow{3}{*}{LR} & \multirow{3}{*}{\begin{tabular}[c]{@{}c@{}}Target\\ Test\\Acc\end{tabular}} & \multirow{3}{*}{\begin{tabular}[c]{@{}c@{}}Target \\Train\\Acc\end{tabular}} &\multicolumn{2}{c}{MIA with Highest Acc}\\ 
% \cline{3-11} 
\cmidrule(lr){6-7} 
\addlinespace 
 & & & & & MIA & Acc \\ 
%\multirow{2}{*}{fire rates} & \multirow{2}{*}{loss} & \multirow{2}{*}{\begin{tabular}[c]{@{}c@{}}prediction \\correctness\end{tabular}} & \multirow{2}{*}{\begin{tabular}[c]{@{}c@{}}top-3 \\fire rates\end{tabular}} & \multirow{2}{*}{\begin{tabular}[c]{@{}c@{}}maximum \\fire rate\end{tabular}} & \multirow{2}{*}{\begin{tabular}[c]{@{}c@{}}logit-scaled \\fire rate\end{tabular}} & \multirow{2}{*}{\begin{tabular}[c]{@{}c@{}}Mentr with \\fire rates\end{tabular}} & \multirow{2}{*}{\begin{tabular}[c]{@{}c@{}}avg membrane\\ potential\end{tabular}}\\ 
% fire rates   loss   prediction \\correctness  top-3 \\fire rates  maximum \\fire rate  logit-scaled \\fire rate  Mentr with \\fire rates avg membrane\\ potential
% (1) 85.8 (2) 88.8 (3) 67.5 (4) 87.6 (5)86.0 (6)87.1 (7)87.4 (8) 50.7  
% (1) 50.0 (2) 63.6 (3)62.6 (4) 61.5 (5)62.8 (6)63.2 (7) 63.6 (8) 50.0
\addlinespace
\hline
\addlinespace 
\multirow{4}{*}{CIFAR10-DVS} & Adam & 0.001 & 61.8\% & 99.3\% & \begin{tabular}[c]{@{}c@{}}Mentr with \\fire rates\end{tabular} & 83.9\%\\
% & TrainInversion & 336   & 48   & 336  & 48  & 336  & 48  & 336  & 48  &    \\
\addlinespace 
& Adam & 0.1 & 10.6\%& 9.8\%& loss& 50.0\%\\
% & TrainInversion & 336   & 48   & 336  & 48  & 336  & 48  & 336  & 48  &    \\
\addlinespace 
& SGD & 0.001 & 10.0\% & 9.6\%& loss& 50.0\%\\
% & TrainInversion & 336   & 48   & 336  & 48  & 336  & 48  & 336  & 48  &    \\
\addlinespace 
& SGD & 0.1 & 65.2\% & 99.9\% & loss & 88.8\%\\
% & TrainInversion & 336   & 48   & 336  & 48  & 336  & 48  & 336  & 48  &    \\
\hline

\addlinespace 
\multirow{4}{*}{N-Caltech101} & Adam & 0.001 & 44.6\% & 80.0\% & loss & 79.5\%\\
% & TrainInversion & 336   & 48   & 336  & 48  & 336  & 48  & 336  & 48  &    \\
\addlinespace 
& Adam & 0.1 & 9.2\%& 1.8\%& loss & 50.0\%\\
% & TrainInversion & 336   & 48   & 336  & 48  & 336  & 48  & 336  & 48  &    \\
\addlinespace 
& SGD & 0.001 & 9.2\% & 4.2\%& loss & 50.0\%\\
% & TrainInversion & 336   & 48   & 336  & 48  & 336  & 48  & 336  & 48  &    \\
\addlinespace 
& SGD & 0.1 & 63.4\% & 87.2\% & loss & 63.6\%\\
\hline

\end{tabular}
\end{adjustbox}
\end{table}

\subsubsection{Varying the Number of Time Steps}

The number of time steps is an important factor for SNNs, as each data point obtains the final prediction result based on the fire rate of the last spiking neurons during those steps. For each step, SNN takes one frame from a data point and updates the membrane potential of its spiking neurons to incur final spikes. We explore the impact of the number of time steps on the accuracy of target models and MIAs as shown in Table~\ref{vary_time_step}. We observe that the gap between the highest attack accuracy obtained with time steps 8 and 16 is within 1\%. For N-Caltech101, the attack accuracy under 4 time steps is higher due to a larger generalization gap. If we increase the number of time steps from 4 to 8, the highest attack accuracy will increase for N-MNIST and CIFAR10-DVS, even though the generalization gap slightly decreases with larger time steps. Hence, increasing the number of time steps would usually slightly increase the vulnerability towards MIAs to a certain stable level. We postulate the main reason for this phenomenon is that increasing the time steps means more frames of a data point will go through the model for updating its parameters, making the model have deeper memorization of the data point and, thus, MIA easier.

\begin{table}[ht]
\centering
\caption{The attack accuracy with various time steps while training SNNs (CNN).}
\label{vary_time_step}
\begin{adjustbox}{max width=0.46\textwidth}
\begin{tabular}{cccccc}
\hline
\addlinespace 
\multirow{3}{*}{\begin{tabular}[c]{@{}c@{}}Dataset\end{tabular}} & \multirow{3}{*}{\begin{tabular}[c]{@{}c@{}}The Number of \\ Time Steps\end{tabular}} & \multirow{3}{*}{\begin{tabular}[c]{@{}c@{}}Target\\ Test\\Acc\end{tabular}} & \multirow{3}{*}{\begin{tabular}[c]{@{}c@{}}Target \\Train\\Acc\end{tabular}} &\multicolumn{2}{c}{MIA with Highest Acc}\\ 
% \cline{3-11} 
\cmidrule(lr){5-6} 
\addlinespace 
 & & & & MIA & Acc \\ 
%\multirow{2}{*}{fire rates} & \multirow{2}{*}{loss} & \multirow{2}{*}{\begin{tabular}[c]{@{}c@{}}prediction \\correctness\end{tabular}} & \multirow{2}{*}{\begin{tabular}[c]{@{}c@{}}top-3 \\fire rates\end{tabular}} & \multirow{2}{*}{\begin{tabular}[c]{@{}c@{}}maximum \\fire rate\end{tabular}} & \multirow{2}{*}{\begin{tabular}[c]{@{}c@{}}logit-scaled \\fire rate\end{tabular}} & \multirow{2}{*}{\begin{tabular}[c]{@{}c@{}}Mentr with \\fire rates\end{tabular}} & \multirow{2}{*}{\begin{tabular}[c]{@{}c@{}}avg membrane\\ potential\end{tabular}}\\ 
% fire rates   loss   prediction \\correctness  top-3 \\fire rates  maximum \\fire rate  logit-scaled \\fire rate  Mentr with \\fire rates avg membrane\\ potential
% (1) 75.0 (2) 79.5 (3) 68.6 (4) 76.4 (5)69.6 (6)80.3 (7)79.5 (8) 71.7 
% (1) 81.2 (2) 83.7 (3) 67.4 (4) 82.0 (5)80.3 (6)83.7 (7)83.7 (8) 71.9
% (1) 70.2 (2) 79.5 (3) 69.4 (4) 78.8 (5)77.9 (6)79.5 (7)79.5 (8) 52.9
% (1) 72.7 (2) 78.0 (3) 67.9 (4) 76.7 (5)77.0 (6)78.0 (7)78.0 (8) 50.2 
% (1) 51.0 (2) 51.4 (3) 50.6 (4) 49.6 (5)51.3 (6)51.7 (7)51.4 (8) 46.8 
% (1) 51.5 (2) 52.3 (3) 50.5 (4) 49.7 (5)51.8 (6)52.3 (7)52.3 (8) 48.4 
\addlinespace
\hline

\addlinespace 
\multirow{3}{*}{N-MNIST} & 4 & 98.7\% & 99.6\% & loss & 51.4\%\\
% & TrainInversion & 336   & 48   & 336  & 48  & 336  & 48  & 336  & 48  &    \\
\addlinespace 
& 8 & 99.0\%& 99.9\%& loss & 52.3\%\\
% & TrainInversion & 336   & 48   & 336  & 48  & 336  & 48  & 336  & 48  &    \\
\addlinespace 
& 16 & 99.2\% & 99.9\%& loss & 52.7\%\\
\hline

\addlinespace 
\multirow{3}{*}{CIFAR10-DVS} & 4 & 62.8\% & 98.1\% & \begin{tabular}[c]{@{}c@{}}Mentr with \\fire rates\end{tabular} & 80.3\%\\
% & TrainInversion & 336   & 48   & 336  & 48  & 336  & 48  & 336  & 48  &    \\
\addlinespace 
& 8 & 65.4\%& 99.3\%& loss& 83.7\%\\
% & TrainInversion & 336   & 48   & 336  & 48  & 336  & 48  & 336  & 48  &    \\
\addlinespace 
& 16 & 64.4\% & 99.5\%& loss& 83.6\%\\
% & TrainInversion & 336   & 48   & 336  & 48  & 336  & 48  & 336  & 48  &    \\
\hline

\addlinespace 
\multirow{3}{*}{N-Caltech101} & 4 & 61.4\% & 99.2\% & loss & 79.5\%\\
% & TrainInversion & 336   & 48   & 336  & 48  & 336  & 48  & 336  & 48  &    \\
\addlinespace 
& 8 & 63.9\%& 99.6\%& loss & 78.0\%\\
% & TrainInversion & 336   & 48   & 336  & 48  & 336  & 48  & 336  & 48  &    \\
\addlinespace 
& 16 & 68.0\% & 99.6\%& \begin{tabular}[c]{@{}c@{}}logit-scaled \\fire rate\end{tabular} & 78.9\%\\
\hline

\end{tabular}
\end{adjustbox}
\end{table}

% \subsection{Ablation Study}

% (1) the surrogate activation function

% (2) the encoding method:  rate encoding, time-to-first-spike encoding

% (3) 

\subsection{Defense Evaluation}
\label{defense}

% (1) Differential Privacy~\cite{wang2022dpsnn} 
%     we can remove the differential privacy if we find enough defense strategies.

Even though reducing the performance of MIA, Differential Privacy (DP) also causes a large utility loss, as demonstrated in the case of  ANNs~\cite{shokri_membership_2017,carlini_membership_2021}. Hence, we apply strategies for improving the generalization of SNNs to defend against MIAs.
Indeed, the generalization gap is a commonly acceptable cause for MIAs~\cite{yeom_privacy_2018}, and the initial works~\cite{shokri_membership_2017,salem_ml-leaks_2018} also attempted to defend against MIAs with regularization and model ensemble. 
For neuromorphic datasets, we apply two strategies for data augmentation. The first strategy is EventDrop~\cite{gu2021eventdrop} to drop events in 1) a time interval, 2) an area of the coordinate, or 3) randomly. Following~\cite{gu2021eventdrop}, we augment each data sample (a group of events) with the three mentioned dropping methods or do not augment them. The second strategy applies geometric augmentations (NDA~\cite{Li_NDA_2022}), including rolling, rotation, cutout~\cite{devries2017improved}, shear, flip, and CutMix~\cite{yun2019cutmix}. Following the mechanism of ``M1N2'' in the original paper, we apply flip and CutMix and randomly select one of the four left augmentation methods for each data sample. For static datasets, we apply basic strategies, including horizontal flip, random crop, and resize for data augmentation. 

Tables~\ref{static_dataset_with_augmentation} and~\ref{neuromorphic_dataset_with_augmentation} show the accuracy change of the target model and MIAs before and after applying data augmentation to static and neuromorphic datasets. From the two tables, we observe that data augmentation will reduce the target model's generalization gap, leading to the performance drop of MIAs. For static datasets, SNNs trained with backpropagation reduce their generalization gap to a value close to 0 with data augmentation, leading to MIAs' performance close to random guessing. 
Meantime, the test accuracy of the target model reduces 10\% to 20\%. The conversion of ANNs trained with data augmentation to SNNs decreases the accuracy of MIAs by about 10\%, with the reduction of the test accuracy by more than 10\%. Notably, the attack accuracy could also be as high as 66.4\% while converting ANNs to SNNs. For ANNs trained with static datasets, the reduced accuracy of MIAs is still about 60\%, which is 10\% to 20\% lower than the one without data augmentation, with about a 10\% decrease in the generalization gap. 
The data augmentation for the static datasets prevents the SNNs trained with backpropagation from being attacked. For ANNs and SNNs converted from ANNs trained with static datasets, the data augmentation dramatically reduces the performance of MIAs.
%\todo{this sentence does not make sense}

Considering neuromorphic datasets and SNNs with a structure of VGG11, NDA~\cite{Li_NDA_2022} reduces the generalization gap to 6.9\% for CIFAR10-DVS and 0.1\% for N-Caltech101, which is smaller than the generalization gap with EventDrop~\cite{gu2021eventdrop} (29.5\% for CIFAR10-DVS and 1.8\% for N-Caltech101). Consequently, the performance of MIAs on SNNs trained with the NDA is lower than on SNNs trained with the EventDrop. For example, the attack accuracy of MIAs on the SNN (VGG11) drops from 83.9\% to 58.3\% while being trained with CIFAR10-DVS augmented with NDA. The attack accuracy is still 75.8\% for the SNN (VGG11) with EventDrop. This indicates NDA is better than EventDrop in reducing the performance of MIAs. Besides, NDA maintains the test accuracy of SNN (VGG11) at 61.4\%, which is slightly higher than the test accuracy of SNN (VGG11) with EventDrop (60.8\%). For N-Caltech101, NDA reduces the performance of SNN (VGG11) to 53.8\%, which is lower than 55.2\% with EventDrop. The test accuracy of SNN (VGG11) is 35.2\% and 32.0\% with EventDrop and NDA. From the above analysis, we can observe that NDA is better than EventDrop in reducing the performance of MIAs. We believe that the geometric augmentations from NDA make the augmented data further from the original data, compared with the augmented data generated via randomly dropping events (EventDrop). Hence, NDA largely increases the generalization of the SNN and reduces the performance of MIAs.

For ANN (VGG11) trained with neuromorphic data, the performance of MIAs slightly increases even though the generation gap is smaller when the augmentation method is EventDrop. For example, the performance of MIA on ANN (VGG11) trained with CIFAR10-DVS increases from 83.8\% to 86.5\% after the application of EventDrop. While training the ANN on neuromorphic data, we accumulate events belonging to a data point into multiple frames and label each frame with the same class for training the ANN. Frames belonging to the same data point are similar as they describe the same object. The dropping of events reduces the dissimilarity between frames from the same data point, which leads to more similar frames. Those similar frames increase the memorization of the target model during training, causing the improvement of MIA. For the NDA augmentation method, the geometric augmentations make previous similar frames different and improve the target model's generalization and test accuracy, reducing the performance of MIAs. For example, the accuracy of MIA drops from 83.8\% to 71.8\% with NDA, and the test accuracy increases from 55.0\% to 63.7\%. The above analysis further indicates that NDA is better at reducing the performance of MIA and even increasing the test accuracy of ANNs trained with neuromorphic data.

Basic augmentation on static data prevents SNNs trained with backpropagation from being attacked and reduces the performance of MIA on the SNNs converted from ANNs. Two augmentation methods on neuromorphic data decrease the performance of MIA on SNNs but cannot completely prevent them as the attack accuracy could also be 58.3\% even for the effective augmentation mechanism, NDA. Moreover, the augmentation could lead to a drop in the test accuracy of the target model, reducing its utility. Therefore, the method to maintain the test accuracy and reduce the performance of MIAs could be an interesting future work.

% Particularly, the NDA keeps the test accuracy of SNN (VGG11) trained with CIFAR10-DVS at about 61\% and largely decreases the training accuracy to 68.3\%, reducing the highest attack accuracy from 83.9\% to 58.3\% (a reduction of 25.6\%). Hence, NDA is better than EventDrop in reducing the performance of MIAs and keeping the test accuracy of the target model.\todo{why?} 
% From the performance of MIAs, we also find those two augmentations cannot completely defend against MIAs.

\begin{table*}[ht]
\centering
\caption{The accuracy change while applying augmentation to static datasets.}
\label{static_dataset_with_augmentation}
\begin{adjustbox}{max width=0.82\textwidth}
\begin{tabular}{ccccccccccc}
\hline
\addlinespace 
\multirow{3}{*}{\begin{tabular}[c]{@{}c@{}}Dataset\end{tabular}} & \multirow{3}{*}{Model} & \multirow{3}{*}{\begin{tabular}[c]{@{}c@{}}Training \\ Strategy\end{tabular}} & \multicolumn{4}{c}{Without Augmentation} & \multicolumn{4}{c}{With Augmentation}\\ 

% \hline
\cmidrule(lr){4-7} \cmidrule(lr){8-11}
% \addlinespace 
 &  & & \multirow{2}{*}{\begin{tabular}[c]{@{}c@{}}Target Test\\Acc\end{tabular}} & \multirow{2}{*}{\begin{tabular}[c]{@{}c@{}}Target Train\\Acc\end{tabular}} &\multicolumn{2}{c}{MIA with Highest Acc} & \multirow{2}{*}{\begin{tabular}[c]{@{}c@{}}Target Test\\Acc\end{tabular}} & \multirow{2}{*}{\begin{tabular}[c]{@{}c@{}}Target Train\\Acc\end{tabular}} &\multicolumn{2}{c}{MIA with Highest Acc}\\ 
% \cline{3-11} 
\cmidrule(lr){6-7} \cmidrule(lr){10-11}
% \addlinespace 
 & & & & & MIA & Acc & & & MIA & Acc\\ 
\hline
\addlinespace 
\multirow{6}{*}{CIFAR-10} & ANN (VGG11) & BP & 80.5\% & 100.0\% & \begin{tabular}[c]{@{}c@{}}Mentr with \\confidence scores\end{tabular} & 70.3\% & 87.6\% & 98.6\% & loss & 59.5\% \\
% 
% & TrainInversion & 336   & 48   & 336  & 48  & 336  & 48  & 336  & 48  &    \\
\addlinespace 
& SNN (VGG11) & BP & 78.3\%& 99.6\%& loss& 69.5\% & 67.3\%& 67.6\%& \begin{tabular}[c]{@{}c@{}}logit-scaled \\fire rate\end{tabular}& 51.7\%\\
% & TrainInversion & 336   & 48   & 336  & 48  & 336  & 48  & 336  & 48  &    \\
\addlinespace 
& SNN (VGG11) & Conversion & 77.0\% & 97.4\%& \begin{tabular}[c]{@{}c@{}}logit-scaled \\fire rate\end{tabular} & 63.1\% & 63.6\% & 70.5\%& \begin{tabular}[c]{@{}c@{}}prediction \\correctness\end{tabular} & 53.3\%\\
% & TrainInversion & 336   & 48   & 336  & 48  & 336  & 48  & 336  & 48  &    \\
\hline

\addlinespace 
\multirow{6}{*}{Caltech101} & ANN (VGG11) & BP & 61.8\% & 99.9\% & \begin{tabular}[c]{@{}c@{}}logit-scaled \\confidence\end{tabular} & 80.2\% & 62.9\% & 91.9\% & \begin{tabular}[c]{@{}c@{}}prediction \\correctness\end{tabular} & 60.5\%\\
% & TrainInversion & 336   & 48   & 336  & 48  & 336  & 48  & 336  & 48  &    \\
\addlinespace 
& SNN (VGG11) & BP & 47.5\%& 45.1\%& \begin{tabular}[c]{@{}c@{}}maximum \\fire rate\end{tabular} & 52.3\% & 30.4\%& 33.0\%& \begin{tabular}[c]{@{}c@{}}avg membrane\\ potential\end{tabular} & 51.7\%\\
% & TrainInversion & 336   & 48   & 336  & 48  & 336  & 48  & 336  & 48  &    \\
\addlinespace 
& SNN (VGG11) & Conversion & 55.5\% & 99.1\%& loss & 72.9\% & 44.0\% & 70.6\%& loss & 66.4\% \\
% & TrainInversion & 336   & 48   & 336  & 48  & 336  & 48  & 336  & 48  &    \\
\hline

\end{tabular}
\end{adjustbox}
\end{table*}

\begin{table*}[ht]
\centering
\caption{The accuracy change while applying augmentation to neuromorphic datasets.}
\label{neuromorphic_dataset_with_augmentation}
\begin{adjustbox}{max width=0.98\textwidth}
\begin{tabular}{cccccccccccccc}
\hline
\addlinespace 
\multirow{3}{*}{\begin{tabular}[c]{@{}c@{}}Dataset\end{tabular}} & \multirow{3}{*}{Model} & \multicolumn{4}{c}{Without Augmentation} & \multicolumn{4}{c}{With EventDrop~\cite{gu2021eventdrop}} & \multicolumn{4}{c}{With NDA~\cite{Li_NDA_2022}}\\ 

% \hline
\cmidrule(lr){3-6} \cmidrule(lr){7-10} \cmidrule(lr){11-14}
% \addlinespace 
 &  & \multirow{2}{*}{\begin{tabular}[c]{@{}c@{}}Target Test\\Acc\end{tabular}} & \multirow{2}{*}{\begin{tabular}[c]{@{}c@{}}Target Train\\Acc\end{tabular}} &\multicolumn{2}{c}{MIA with Highest Acc} & \multirow{2}{*}{\begin{tabular}[c]{@{}c@{}}Target Test\\Acc\end{tabular}} & \multirow{2}{*}{\begin{tabular}[c]{@{}c@{}}Target Train\\Acc\end{tabular}} &\multicolumn{2}{c}{MIA with Highest Acc} & \multirow{2}{*}{\begin{tabular}[c]{@{}c@{}}Target Test\\Acc\end{tabular}} & \multirow{2}{*}{\begin{tabular}[c]{@{}c@{}}Target Train\\Acc\end{tabular}} &\multicolumn{2}{c}{MIA with Highest Acc}\\ 
% \cline{3-11} 
\cmidrule(lr){5-6} \cmidrule(lr){9-10} \cmidrule(lr){13-14}
% \addlinespace 
 & & & & MIA & Acc & & & MIA & Acc & & & MIA & Acc \\ 
\hline
\addlinespace 
\multirow{4}{*}{CIFAR10-DVS} & ANN (VGG11) & 55.0\% & 100.0\% & hinge loss & 83.8\%  & 55.1\% & 89.3\% & \begin{tabular}[c]{@{}c@{}}logit-scaled \\confidence\end{tabular} & 86.5\% & 63.7\% & 95.1\% & \begin{tabular}[c]{@{}c@{}}Mentr with \\confidence scores\end{tabular} & 71.8\% \\
% & TrainInversion & 336   & 48   & 336  & 48  & 336  & 48  & 336  & 48  &    \\
\addlinespace 
& SNN (VGG11) & 61.8\% & 99.3\% & \begin{tabular}[c]{@{}c@{}}Mentr with \\fire rates\end{tabular} & 83.9\%  & 60.8\% & 90.3\% & \begin{tabular}[c]{@{}c@{}}Mentr with \\fire rates\end{tabular} & 75.6\% & 61.4\% & 68.3\% & \begin{tabular}[c]{@{}c@{}}Mentr with \\fire rates\end{tabular} & 58.3\%\\
% & TrainInversion & 336   & 48   & 336  & 48  & 336  & 48  & 336  & 48  &    \\
\hline

\addlinespace 
\multirow{4}{*}{N-Caltech101} & ANN (VGG11) & 56.9\% & 100.0\% & loss & 78.6\% & 55.8\% & 88.7\% & hinge loss & 79.4\% & 60.7\% & 99.4\% & loss & 76.9\%\\
% & TrainInversion & 336   & 48   & 336  & 48  & 336  & 48  & 336  & 48  &    \\
\addlinespace  
& SNN (VGG11) & 44.6\% & 80.0\%& loss & 79.5\%& 35.2\% & 37.0\%& loss & 55.2\% & 32.0\% & 32.1\%& \begin{tabular}[c]{@{}c@{}}top-3 \\fire rates\end{tabular} & 53.8\%\\
% & TrainInversion & 336   & 48   & 336  & 48  & 336  & 48  & 336  & 48  &    \\
\hline

\end{tabular}
\end{adjustbox}
\end{table*}

%The first strategy is to add a voting layer with more neurons as the final prediction layer~\cite{Fan_2021_Learnable_Membrane_Time_Constant}.
% (2) A method to improve the robustness of the SNN might be a solution for defending
%     - insert a voting layer after the last~\cite{Fan_2021_Learnable_Membrane_Time_Constant}
%     - eventDrop~\cite{gu2021eventdrop}
%     - geometric augmentations~\cite{Li_NDA_2022}

%% file: sections/related_works.tex
\section{Related Works}
\label{sec:related}

This part introduces previous works on MIA and attacks against SNNs. 

\subsection{Membership Inference Attack}
\label{related_work_MIA}

As people and governments pay more attention to personal data privacy, the development and research on MIA are flourishing, as discussed next.
Since the successful membership inference attack in 2017~\cite{shokri_membership_2017}, the following works put much effort into ways to improve the performance of MIAs~\cite{salem_ml-leaks_2018,yeom_privacy_2018,carlini_membership_2021,Ye_Enhanced_2022}, attacking strategies with less information~\cite{li_label-leaks_2020,choquette-choo_label-only_2021,Hui_21_BlindMIA,label_only_MIA_Mauro}, MIAs on various datasets or models~\cite{hayes_logan_2019,kong2023efficient}, vulnerability of data points~\cite{yaghini_disparate_2019,conti2022vulnerability}, and defenses against MIAs~\cite{nasr_machine_2018,jia_memguard_2019,chen2023overconfidence}. 

Shokri et al.~\cite{shokri_membership_2017} proposed training multiple shadow models to mimic the behavior of the target model and extracting a dataset from the shadow model and its training and test data to train the attack model. Salem et al.~\cite{salem_ml-leaks_2018} relaxed assumptions on the number of shadow models and the auxiliary data. Yeom et al.~\cite{yeom_privacy_2018} proposed to utilize loss for MIA, and Ye et al.~\cite{Ye_Enhanced_2022} enhanced MIAs based on loss with carefully designed thresholds. 

Li et al.~\cite{li_label-leaks_2020}, and Choquette-choo et al.~\cite{choquette-choo_label-only_2021} leveraged the distance between the original data point and its black-box adversarial example as the signal of MIA. Hui et al.~\cite{Hui_21_BlindMIA} eliminated shadow models by measuring the distance change between the evaluation and non-member data to implement an MIA. Hayes et al.~\cite{hayes_logan_2019} investigated MIA on GANs, while Kong et al.~\cite{kong2023efficient} implemented MIA on the diffusion model. Yaghini et al.~\cite{yaghini_disparate_2019} explored the vulnerability of a subgroup under the MIA. In contrast, Conti et al.~\cite{conti2022vulnerability} investigated the vulnerability of one data point under multiple target models and MIAs. Nasr et al.~\cite{nasr_machine_2018} explored defenses against MIA with the adversarial training algorithm, while Jia et al.~\cite{jia_memguard_2019} proposed to find adversarial examples of the attack model.

\subsection{The Attacks against Spiking Neural Network}
\label{related_work_SNN}

Works explored adversarial examples and backdoor attacks against SNNs. Abad et al.~\cite{abad2024sneaky} systematically investigated the backdoor attack against the SNN. Specifically, they explored the trigger position, polarity, and size while inserting the same static trigger for all frames of a data point. To improve the moving backdoor in previous work~\cite{Abad2022Backdoor}, the authors conducted a complete experimental setup to find the best moving trigger, which changes the location over different frames. Besides, they proposed a smart backdoor to detect the most active region and insert moving triggers with the least used polarity in this region. Finally, they investigated dynamic moving backdoors where the triggers are invisible and unique for each image and frame via optimizing a spiking autoencoder to generate noise during SNN training. 

Nomura et al.~\cite{Nomura2022IEEE} explored the robustness of the time-to-first-spike encoding SNNs against white-box adversarial examples. Their result indicated the time-to-first-spike encoding SNNs are more robust than ANNs when trained with an appropriate temporal penalty setting, specifying the output spikes' timing to approximate the reference timing. Sharmin et al.~\cite{sharmin2019comprehensive} systematically analyzed the adversarial robustness of SNNs and concluded the dependence of robustness on the SNN training mechanism. 

To apply a gradient-based adversarial attack on SNNs trained with backpropagation and a surrogate activation neuron, Liang et al.~\cite{liang2021exploring} proposed two strategies to solve the problem of gradient-input incompatibility and gradient vanishing. Specifically, they designed a gradient-to-spike (G2S) converter to convert continuous gradients to ternary ones compatible with spike inputs. Then, they proposed a restricted spike flipper (RSF) to construct ternary gradients that can randomly flip the spike inputs when facing all-zero gradient maps, where the turnover rate of inputs is controllable.

%% file: sections/conclusion.tex
\section{Conclusions and Future Work}
\label{sec:conclusions}

This work evaluates the membership privacy leakage of SNNs with eight MIAs and compares the vulnerability of SNNs and ANNs when facing MIAs. Our results show that SNNs suffer from MIAs, especially when trained with neuromorphic datasets, where SNNs are more vulnerable than ANNs. 
If we convert ANNs to SNNs, the accuracy of MIAs will drop with a relatively low reduction in the accuracy of the original classification task. Hence, we recommend converting from ANNs to SNNs to reduce the leakage of membership privacy. 
The rule that a higher generalization gap usually leads to a higher performance of MIAs is also applicable to SNNs from our exploration of various datasets, model types, and the hyperparameter study.
Moreover, the basic data augmentation for static datasets and two recent data augmentation methods for neuromorphic datasets can improve the generalization of SNNs and eliminate MIAs. However, data augmentations cannot completely prevent MIAs. As the training accuracy of SNNs with data augmentation is usually lower than those without augmentation, we leave the method to reduce the generalization gap and keep the training accuracy of SNNs for future work. Finally, considering that SNNs are vulnerable to MIAs, we plan to investigate further on possible defense mechanisms.

% Moreover, we add synthetic data to neuromorphic datasets from two data augmentations to improve the generalization of SNNs and eliminate MIAs. Between the two data augmentation mechanisms, we observe that the NDA keeps the test accuracy of one SNN and simultaneously reduces the highest attack accuracy of MIAs, which is more suitable than the EventDrop. As the training accuracy of SNNs with data augmentation is usually lower than those without augmentation, we leave the method to reduce the generalization gap and keep the training accuracy of SNNs for future work.

% (within 9\%)

%% file: sections/appendix.tex
%\section{Additional Experimental Results}

\subsection{MIAs against SNNs}
\label{MIAs_against_SNNs}

Table~\ref{MIAs_SNNs} provides the attack accuracy of eight MIAs against SNNs trained with various model types and datasets. The training strategy ``bp'' means backpropagation, and ``conversion'' means converting ANNs to obtain SNNs. ``Target Test Acc'' and ``Target Train Acc'' represent the test and training accuracy of the target model. Note that the accuracy of the shadow model is close to the accuracy of the target model as the shadow model mimics the target model. 
We explain eight MIAs and how to apply them in Section~\ref{evaluation_methodology}. 

\begin{table*}[ht]
\centering
\caption{The attack accuracy of MIAs against SNNs.}
\label{MIAs_SNNs}
\begin{adjustbox}{max width=0.98\textwidth}
\begin{tabular}{ccccccccccccc}
\hline
\addlinespace 
\multirow{3}{*}{Dataset} & \multirow{3}{*}{\begin{tabular}[c]{@{}c@{}}Model \\Type\end{tabular}} &\multirow{3}{*}{\begin{tabular}[c]{@{}c@{}}Training \\Strategy\end{tabular}} & \multirow{3}{*}{\begin{tabular}[c]{@{}c@{}}Target Test\\Acc\end{tabular}} & \multirow{3}{*}{\begin{tabular}[c]{@{}c@{}}Target Train\\Acc\end{tabular}} &\multicolumn{8}{c}{Attack Accuracy of MIAs}\\ 
% \cline{3-11} 
\cmidrule(lr){6-13} 
\addlinespace 
 & & & & & \multirow{2}{*}{fire rates} & \multirow{2}{*}{loss} & \multirow{2}{*}{\begin{tabular}[c]{@{}c@{}}prediction \\correctness\end{tabular}} & \multirow{2}{*}{\begin{tabular}[c]{@{}c@{}}top-3 \\fire rates\end{tabular}} & \multirow{2}{*}{\begin{tabular}[c]{@{}c@{}}maximum \\fire rate\end{tabular}} & \multirow{2}{*}{\begin{tabular}[c]{@{}c@{}}logit-scaled \\fire rate\end{tabular}} & \multirow{2}{*}{\begin{tabular}[c]{@{}c@{}}Mentr with \\fire rates\end{tabular}} & \multirow{2}{*}{\begin{tabular}[c]{@{}c@{}}avg membrane\\ potential\end{tabular}}\\ 
&&&&&\\
\hline
\addlinespace 
MNIST & SNN (CNN) & Bp & 99.4\% & 99.9\% & 50.2\% & 51.0\%& 50.3\%& 50.0\% & 50.7\%& 51.0\%& 51.0\%& 50.7\%\\
\addlinespace  
N-MNIST & SNN (CNN) & Bp & 99.2\% & 99.9\%& 51.0\% & 52.7\%& 50.4\%& 49.3\% & 52.4\%& 52.7\%& 52.7\%& 48.1\%\\
\addlinespace
MNIST & SNN (CNN) & Conversion & 99.4\% & 99.9\%& 50.0\% & 51.2\%& 50.4\%& 50.8\% & 51.1\%& 51.2\%& 51.2\%&50.2\%\\
\addlinespace 
MNIST & SNN (VGG11) & Bp & 99.3\% & 99.9\% & 50.2\% & 51.1\%& 50.3\%& 50.9\% & 50.8\%& 51.1\%& 51.2\%& 50.1\%\\
\addlinespace  
N-MNIST & SNN (VGG11) & Bp & 99.0\% & 99.9\% & 50.5\% & 51.9\%& 50.5\%& 49.3\% & 49.4\%& 49.6\%& 52.0\%& 50.0\%\\
\addlinespace
MNIST & SNN (VGG11) & Conversion & 99.1\% & 99.9\% & 49.7\% & 50.5\%& 50.5\%& 50.2\% & 50.1\%& 50.1\%& 50.5\%& 49.5\%\\
\addlinespace 
CIFAR-10 & SNN (CNN) & Bp & 78.9\% & 96.6\% & 63.5\% & 65.9\%& 59.8\%& 64.0\% & 63.2\%& 65.9\%& 65.9\%& 59.1\%\\
\addlinespace  
CIFAR10-DVS & SNN (CNN) & Bp & 64.4\% & 99.5\%& 79.7\% & 83.6\%& 67.9\%& 83.1\% & 79.7\%& 82.4\%& 83.6\%& 69.3\%\\
\addlinespace
CIFAR-10 & SNN (CNN) & Conversion & 54.9\% & 66.3\%& 54.3\% & 56.3\%& 56.5\%& 52.2\% & 51.4\%& 56.3\%& 56.4\%&51.6\%\\
\addlinespace 
CIFAR-10 & SNN (VGG11) & Bp & 78.3\% & 99.6\% & 67.4\% & 69.5\%& 60.7\%& 67.8\% & 67.4\%& 69.3\%& 69.5\%& 47.3\%\\
\addlinespace  
CIFAR10-DVS & SNN (VGG11) & Bp & 61.8\% & 99.3\%& 78.9\% & 83.8\%& 69.3\%& 82.5\% & 81.4\%& 82.3\%& 83.9\%& 50.3\%\\
\addlinespace
CIFAR-10 & SNN (VGG11) & Conversion & 77.0\% & 97.4\%& 52.5\% & 61.6\%& 60.4\%& 56.7\% & 58.4\%& 63.1\%& 61.6\%& 49.2\%\\
\addlinespace 
CIFAR-10 & SNN (ResNet18) & Bp & 72.0\% & 98.6\% & 65.7\% & 69.4\%& 63.6\%& 66.2\% & 65.6\%& 69.4\%& 69.4\%& 51.8\%\\
\addlinespace  
CIFAR10-DVS & SNN (ResNet18) & Bp & 49.4\% & 100.0\%& 90.4\% & 87.7\%& 75.4\%& 92.2\% & 91.3\%& 92.8\%& 87.9\%& 60.1\%\\
\addlinespace
CIFAR-10 & SNN (ResNet18) & Conversion & 74.0\% & 94.2\%& 58.2\% & 61.3\%& 59.8\%& 58.8\% & 58.9\%& 61.2\%& 61.4\%& 50.2\%\\
\addlinespace 
Caltech101 & SNN (CNN) & Bp & 71.0\% & 99.9\% & 65.9\% & 74.1\%& 64.4\%& 73.7\% & 73.0\%& 74.1\%& 74.1\%& 56.0\%\\
\addlinespace  
N-Caltech101 & SNN (CNN) & Bp & 68.0\% & 99.6\%& 74.8\% & 78.8\%& 65.8\%& 78.8\% & 78.1\%& 78.9\%& 78.8\%&55.5\%\\
\addlinespace
Caltech101 & SNN (CNN) & Conversion & 51.6\% & 77.4\%& 50.0\% & 67.5\%& 67.8\%& 60.2\% & 62.0\%& 67.6\%& 67.5\%&50.1\%\\
\addlinespace 
Caltech101 & SNN (VGG11) & Bp & 47.5\% & 45.1\% & 50.0\% & 52.0\%& 49.8\%& 50.6\% & 52.3\%& 49.4\%& 49.5\%& 49.3\%\\
\addlinespace  
N-Caltech101 & SNN (VGG11) & Bp & 44.6\% & 80.0\%& 50.0\% & 79.5\%& 70.3\%& 65.2\% & 64.9\%& 70.2\%& 70.2\%& 49.6\%\\
\addlinespace
Caltech101 & SNN (VGG11) & Conversion & 55.5\% & 99.1\%& 56.0\% & 72.9\%& 70.8\%& 70.4\% & 70.7\%& 74.2\%& 72.9\%&50.0\%\\
\addlinespace 
Caltech101 & SNN (ResNet18) & Bp & 65.7\% & 99.9\% & 78.9\% & 78.1\%& 67.2\%& 85.5\% & 85.4\%& 85.0\%& 79.4\%& 53.2\%\\
\addlinespace  
N-Caltech101 & SNN (ResNet18) & Bp & 64.4\% & 100.0\%& 74.8\% & 78.3\%& 67.9\%& 78.6\% & 76.6\%& 75.8\%& 79.4\%& 53.1\%\\
\addlinespace
Caltech101 & SNN (ResNet18) & Conversion & 59.0\% & 99.8\%& 61.2\% & 72.9\%& 69.7\%& 67.4\% & 69.4\%& 72.9\%& 72.9\%&53.4\%\\
\hline
\end{tabular}
\end{adjustbox}
\end{table*}

% (1) confidence scores~\cite{shokri_membership_2017}
% (2) loss~\cite{yeom_privacy_2018}
% (3) prediction correctness~\cite{yeom_privacy_2018}
% (4) top 3 confidence scores~\cite{salem_ml-leaks_2018}
% (5) maximum confidence score~\cite{salem_ml-leaks_2018}
% (6) logit-scaled confidence~\cite{carlini_membership_2021}
% (7) hinge loss~\cite{carlini_membership_2021}
% (8) Mentr with confidence scores~\cite{song_systematic_2021}

\subsection{MIAs against ANNs}
\label{MIAs_against_ANNs}

Table~\ref{MIAs_ANNs} provides the attack accuracy of eight MIAs against ANNs trained with various model types and datasets. The meaning of each column is the same as for Table~\ref{MIAs_SNNs}. %We explain eight MIAs in Section~\ref{evaluation_methodology}.

\begin{table*}[ht]
\centering
\caption{The attack accuracy of MIAs against ANNs.}
\label{MIAs_ANNs}
\begin{adjustbox}{max width=0.98\textwidth}
\begin{tabular}{ccccccccccccc}
\hline
\addlinespace 
\multirow{3}{*}{Dataset} & \multirow{3}{*}{\begin{tabular}[c]{@{}c@{}}Model \\Type\end{tabular}} &\multirow{3}{*}{\begin{tabular}[c]{@{}c@{}}Training \\Strategy\end{tabular}} & \multirow{3}{*}{\begin{tabular}[c]{@{}c@{}}Target Test\\Acc\end{tabular}} & \multirow{3}{*}{\begin{tabular}[c]{@{}c@{}}Target Train\\Acc\end{tabular}} &\multicolumn{8}{c}{Attack Accuracy of MIAs}\\ 
% \cline{3-11} 
\cmidrule(lr){6-13} 
\addlinespace 
 & & & & & \multirow{2}{*}{\begin{tabular}[c]{@{}c@{}}confidence \\scores\end{tabular}} & \multirow{2}{*}{loss} & \multirow{2}{*}{\begin{tabular}[c]{@{}c@{}}prediction \\correctness\end{tabular}} & \multirow{2}{*}{\begin{tabular}[c]{@{}c@{}}top-3 \\confidence scores\end{tabular}} & \multirow{2}{*}{\begin{tabular}[c]{@{}c@{}}maximum \\confidence score\end{tabular}} & \multirow{2}{*}{\begin{tabular}[c]{@{}c@{}}logit-scaled \\confidence\end{tabular}} & \multirow{2}{*}{\begin{tabular}[c]{@{}c@{}}Mentr with \\confidence scores\end{tabular}} & \multirow{2}{*}{hinge loss}\\ 
&&&&&\\
\hline
\addlinespace 
MNIST & ANN (CNN) & Bp & 99.3\%& 99.8\%& 49.5\% & 51.1\%& 50.3\%& 50.8\% & 51.2\%& 51.2\%& 51.2\%&51.2\%\\
\addlinespace 
N-MNIST & ANN (CNN) & Bp & 98.6\%& 99.4\%& 51.3\% & 51.5\%& 50.6\%& 49.1\% & 51.5\%& 51.5\%& 51.4\%&51.6\%\\
\addlinespace 
MNIST & ANN (VGG11) & Bp &  99.4\% & 100.0\% & 50.2\% & 51.0\%& 50.3\%& 50.6\% & 51.0\%& 51.0\%& 51.0\%& 50.8\%\\
\addlinespace 
N-MNIST & ANN (VGG11) & Bp & 98.9\% & 100.0\% & 50.4\% & 51.6\%& 50.5\%& 51.3\% & 51.5\%& 51.6\%& 51.6\%& 51.6\%\\
\addlinespace 
CIFAR-10 & ANN (CNN) & Bp & 80.8\% & 99.4\% & 65.2\% & 68.0\%& 59.6\%& 66.7\% & 66.9\%& 68.0\%& 68.0\%& 68.1\%\\
\addlinespace 
CIFAR10-DVS & ANN (CNN) & Bp & 56.2\%& 87.4\%& 62.2\% & 73.5\%& 71.4\%& 69.7\% & 69.9\%& 73.5\%& 72.9\%&73.5\%\\
\addlinespace 
CIFAR-10 & ANN (VGG11) & Bp & 80.5\% & 100.0\% & 61.2\% & 70.1\%& 60.0\%& 66.5\% & 69.1\%& 70.1\%& 70.3\%& 70.2\%\\
\addlinespace 
CIFAR10-DVS & ANN (VGG11) & Bp & 55.0\%& 100.0\%& 76.6\% & 83.6\%& 72.3\%& 79.6\% & 81.6\%& 83.6\%& 83.4\%&83.8\%\\
\addlinespace
CIFAR-10 & ANN (ResNet18) & Bp & 82.0\% & 100.0\% & 63.9\% & 72.8\%& 59.0\%& 69.3\% & 72.1\%& 72.8\%& 72.9\%& 72.7\%\\
\addlinespace 
CIFAR10-DVS & ANN (ResNet18) & Bp & 57.9\%& 100.0\%& 79.0\% & 84.3\%& 72.3\%& 70.8\% & 82.2\%& 84.3\%& 84.2\%&84.2\%\\
\addlinespace 
Caltech101 & ANN (CNN) & Bp & 71.4\% & 93.9\% & 68.4\% & 69.6\%& 64.2\%& 69.6\% & 69.6\%& 69.7\%& 68.4\%& 69.4\%\\
\addlinespace 
N-Caltech101 & ANN (CNN) & Bp & 61.0\%& 99.2\%& 70.0\% & 80.0\%& 69.6\%& 79.1\% & 79.7\%& 80.0\%& 79.7\%&79.9\%\\
\addlinespace 
Caltech101 & ANN (VGG11) & Bp & 61.8\% & 99.9\% & 58.6\% & 79.6\%& 69.2\%& 78.2\% & 79.4\%& 80.2\%& 79.9\%& 79.5\%\\
\addlinespace 
N-Caltech101 & ANN (VGG11) & Bp & 56.9\%& 100.0\%& 67.7\% & 78.6\%& 71.6\%& 76.1\% & 77.7\%& 78.6\%& 78.6\%&78.5\%\\
\addlinespace 
Caltech101 & ANN (ResNet18) & Bp & 66.1\% & 99.9\% & 50.0\% & 77.5\%& 66.8\%& 78.0\% & 77.1\%& 77.7\%& 77.8\%& 78.5\%\\
\addlinespace 
N-Caltech101 & ANN (ResNet18) & Bp & 59.4\%& 100.0\%& 63.7\% & 79.2\%& 70.4\%& 74.4\% & 78.5\%& 79.2\%& 79.2\%&79.3\%\\

\hline
\end{tabular}
\end{adjustbox}
\end{table*}

\subsection{Accuracy vs. Training Epochs}
\label{accuracy_along_epoch}

\begin{figure*}
    % \Description[Analysis along the training epochs]
    \centering
    \begin{subfigure}[b]{0.98\textwidth}
        \centering
        \includegraphics[scale=0.4]{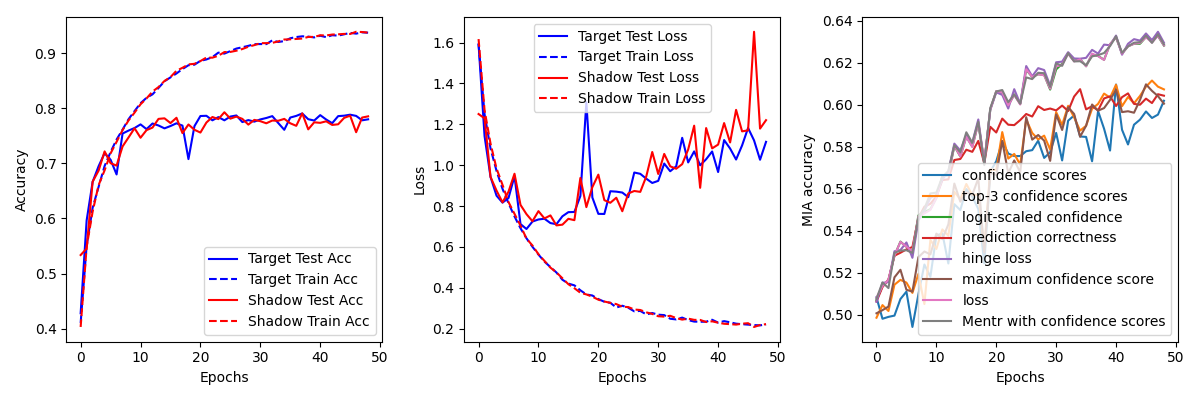}
        \caption{ANN (CNN) and CIFAR-10}
        \label{ANN and CIFAR-10}
    \end{subfigure}
    %  \hfill
    \begin{subfigure}[b]{0.98\textwidth}
        \centering
        \includegraphics[scale=0.4]{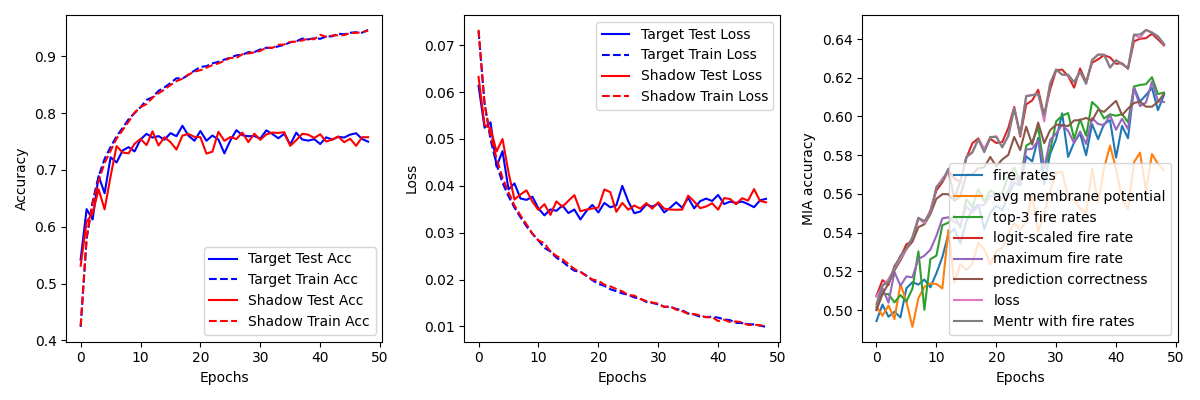}
        \caption{SNN (CNN) and CIFAR-10}
        \label{SNN and CIFAR-10}
    \end{subfigure}
    \begin{subfigure}[b]{0.98\textwidth}
        \centering
        \includegraphics[scale=0.4]{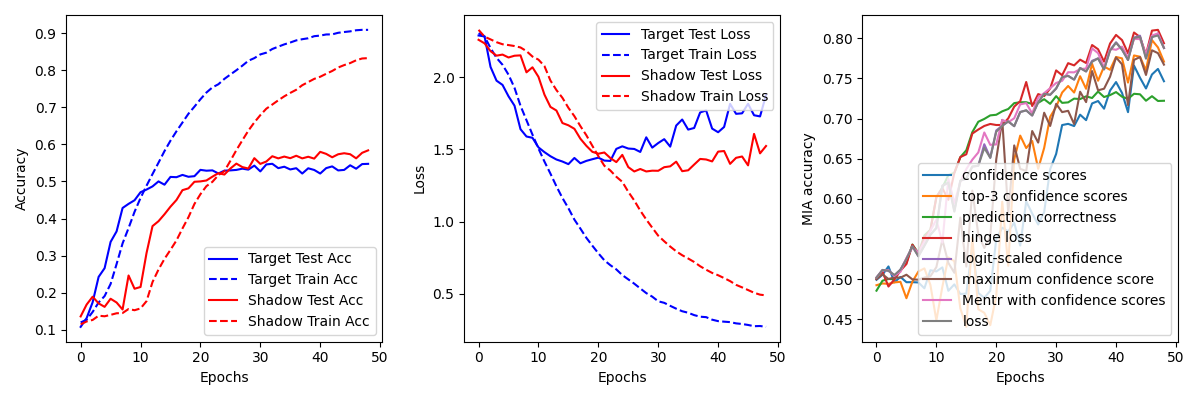}
        \caption{ANN (CNN) and CIFAR10-DVS}
        \label{ANN and CIFAR10-DVS}
    \end{subfigure}
    \begin{subfigure}[b]{0.98\textwidth}
        \centering
        \includegraphics[scale=0.4]{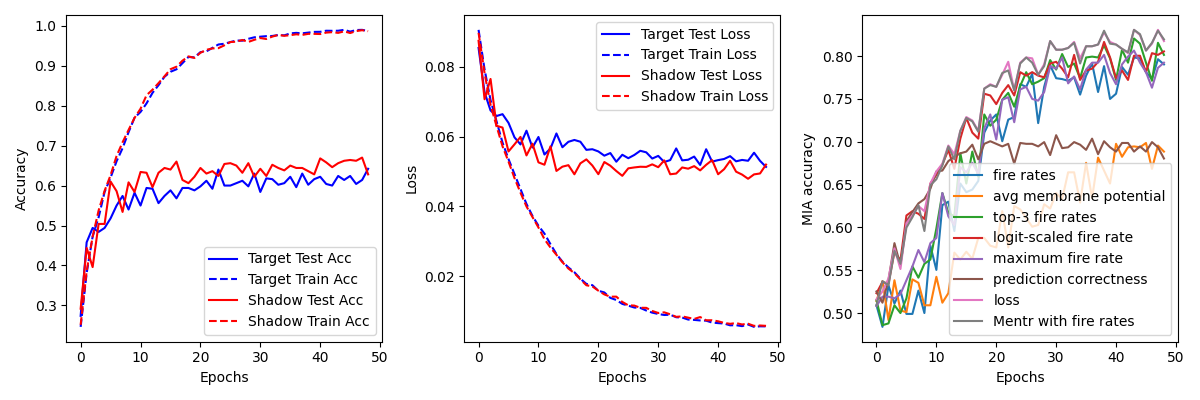}
        \caption{SNN (CNN) and CIFAR10-DVS}
        \label{SNN and CIFAR10-DVS}
    \end{subfigure}
    \caption{The modification of the accuracy of the original task, loss, and attack accuracy of MIAs along the training epochs.}
    \label{modification_along_epochs}
\end{figure*}

% \todo{Does it mean that we can not use Fig if it is at the beginning of the sentence?}
Figure~\ref{modification_along_epochs} shows how the loss, accuracy, and attack accuracy change during the training epochs of the target and shadow models. In the first image of each sub-figure, we draw the change of the training and test accuracy of target and shadow models. Each sub-figure's second image describes the loss modification during the training. The last image of each sub-figure shows the change in attack accuracy within MIAs explored in this work. Each sub-figure is related to a dataset and model type.

